\documentclass[aps,pra,reprint,superscriptaddress,longbibliography]{revtex4-1}

\usepackage{amsmath,amsfonts,bm,graphicx,verbatim,mathrsfs,units}

\usepackage{color,colordvi}
\definecolor{blue}{rgb}{0,0,1}
\definecolor{grey}{rgb}{0.6,0.6,0.6}

\definecolor{myurlcolor}{rgb}{0,0,0.7}
\definecolor{myrefcolor}{rgb}{0.8,0,0}
\usepackage[unicode=true,pdfusetitle, bookmarks=false,bookmarksnumbered=false,
bookmarksopen=false, breaklinks=false,pdfborder={0 0 0},backref=false,
colorlinks=true, linkcolor=myrefcolor,citecolor=myurlcolor,urlcolor=myurlcolor]{hyperref}

\newcommand{\figref}[1]{Fig.~\ref{#1}}
\newcommand{\tabref}[1]{Tab.~\ref{#1}}

\newcommand{\avg}[1]{\left\langle #1 \right\rangle}

\newcommand{\Tr}{\text{Tr}}

\begin{document}


\title{Quantum Thermal Machine as a Thermometer}

\author{Patrick P. Hofer}
\email{patrick.hofer@unige.ch}
\affiliation{D\'epartement de Physique Appliqu\'ee, Universit\'e de Gen\`eve, 1211 Gen\`eve, Switzerland}
\author{Jonatan Bohr Brask}
\affiliation{D\'epartement de Physique Appliqu\'ee, Universit\'e de Gen\`eve, 1211 Gen\`eve, Switzerland}
\author{Mart\' i Perarnau-Llobet}\affiliation{Max-Planck-Institut f\"ur Quantenoptik, Hans-Kopfermann-Str. 1, D-85748 Garching, Germany}\affiliation{ICFO-Institut de Ciencies Fotoniques, The Barcelona Institute of Science and Technology, 08860 Castelldefels, Barcelona, Spain}
\author{Nicolas Brunner}
\affiliation{D\'epartement de Physique Appliqu\'ee, Universit\'e de Gen\`eve, 1211 Gen\`eve, Switzerland}

\date{\today}

\begin{abstract}
We propose the use of a quantum thermal machine for low-temperature thermometry. A hot thermal reservoir coupled to the machine allows for simultaneously cooling the sample while determining its temperature without knowing the model-dependent coupling constants. In its most simple form, the proposed scheme works for all thermal machines which perform at Otto efficiency and can reach Carnot efficiency. We consider a circuit QED implementation which allows for precise thermometry down to $\sim 15\,$mK with realistic parameters. Based on the quantum Fisher information, this is close to the optimal achievable performance. {This implementation demonstrates that our proposal is particularly promising in systems where thermalization between different components of an experimental setup cannot be guaranteed.}
\end{abstract}


\maketitle


\textit{Introduction.---} Accurate sensing and measuring of temperature is of crucial importance throughout natural science and technology. Increased capabilities of control and imaging on smaller and smaller scales have led to the need for precise thermometry down to millikelvin temperatures at sub-micron scales. Conventional techniques are not applicable in this regime, resulting in the development of a broad range of new methods over the last decade \cite{carlospalacio:2016}. Many of these employ probes which are so small that quantum effects become relevant in their design and sensing capabilities, e.g.~quantum dots \cite{walker:2003,seilmeier:2014,haupt:2014}, nitrogen-vacancy centers in diamond \cite{toyli:2013,kucsko:2013,neumann:2013}, superconducting quantum interference devices \cite{halbertal:2016}, and even biomolecules \cite{donner:2012}.
At the same time, the study of thermal processes in the quantum regime has recently seen increased interest fueled by tools developed in quantum information theory \cite{goold:2016,vinjanampathy:2016}. This approach has led to novel insights into the limitations of measuring cold temperatures posed by quantum theory \cite{stace:2010,marzolino:2013,correa:2015,paris:2016,pasquale:2016}, showing that coherence can be beneficial for low temperature thermometry \cite{stace:2010,jevtic:2015,johnson:2016,martin:2013,sabin:2014}.

In a standard approach to thermometry, a probe is brought into thermal contact with the sample and the system is allowed to equilibrate \cite{haupt:2014,seilmeier:2014}. The temperature is then read out through some observable on the probe whose relation to the temperature is known. The measurement can possibly be improved by letting the probe interact with the sample for a finite time only, making use of the transient dynamics \cite{guo:2015,jevtic:2015}, or by increasing the coupling strength between sample and probe \cite{correa:2016}. Both of these approaches lead to a non-equilibrium state for the probe. Another approach to thermometry, which is employed to measure electronic temperatures, makes use of a voltage bias that creates an out-of-equilibrium situation. The temperature can then be determined through the current-voltage characteristics \cite{giazotto:2006,iftikhar:2016,zgirski:2017}. We note that these strategies generally lead to unwanted heating of the sample.

\begin{figure}[t!]
  \centering
  \includegraphics[width=\columnwidth]{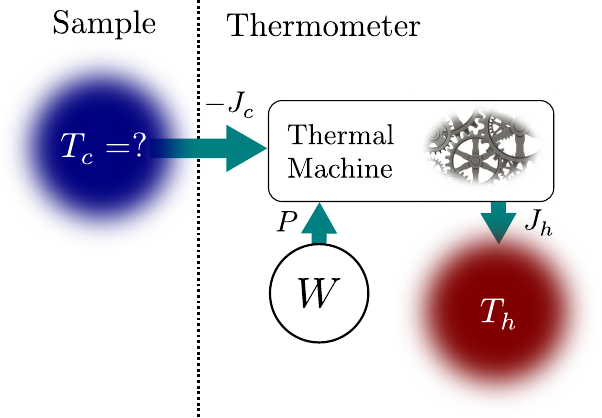}
  \caption{Sketch of the thermometer. In order to measure the temperature of the cold bath, a quantum thermal machine is operated as a refrigerator, inducing a heat current from the cold bath to the hot bath. This requires energy from the work source $W$. The hot temperature is then increased until the machine reaches the Carnot point where its power consumption and the heat flows vanish (i.e. $P=J_c=J_h=0$). For machines that perform with a well known efficiency (e.g. the Otto efficiency $\eta=1-\Omega_c/\Omega_h$), the cold temperature can then be deduced from the hot temperature through the relation $\eta=\eta_C=1-T_c/T_h$, without the need of knowing any model-dependent coupling constants. In this way, a low precision measurement of a hot temperature is converted into a high precision measurement of a low temperature. Operating the thermal machine as a refrigerator avoids any heating of the cold bath.}
  \label{fig:sketch}
\end{figure}

In this letter, we connect thermometry to quantum thermal machines. Such machines are extensively studied to investigate fundamental as well as practical aspects of quantum thermodynamics \cite{goold:2016,benenti:2017,kosloff:2014,quan:2007,brunner:2012}. By construction, these machines constitute out-of-equilibrium systems including a temperature gradient. Here we consider a quantum refrigerator to simultaneously cool the sample and estimate its temperature. This way, the proposed thermometer does not induce any heating of the sample, even if it is at the coldest temperature that is experimentally available. 
{Our proposal thus makes use of a \textit{thermal} bias to create an out-of-equilibrium situation that is favorable for thermometry. This idea goes back to Thomson (Lord Kelvin), who considered the use of a Carnot engine to determine an absolute temperature scale \cite{kelvin:1848} (see also Ref.~\cite{geusic:1967}).}
Note that albeit the thermal bias, the sample is assumed to remain in \textit{local} equilibrium throughout the measurement.

The main idea is illustrated in \figref{fig:sketch}. The sample to be measured is a thermal bath at a cold temperature $T_c$. Through a small quantum system (the machine), the sample is coupled to another bath at a higher temperature $T_h$ and an external power source (which in principle could be provided by a third thermal bath \cite{linden:2010prl,skrzypczyk:2011,hofer:2016,mitchison:2016}). Note that this setup can operate either as a refrigerator, with the power source driving a heat flow from the cold to the hot bath, or as a heat engine, where work is generated using a heat flow from the hot to the cold bath \cite{levy:2012,kosloff:2014,mitchison:2015,silva:2015,brunner:2014,brask:2015,roulet:2017}. Since we are interested in determining $T_c$, the whole setup, apart from the cold bath, should be considered as the thermometer. By operating the machine as a refrigerator, the sample will be cooled during the measurement of $T_c$, avoiding any undesirable heating. Furthermore, by approaching the Carnot point, where the machine approaches reversibility, the need for knowledge of the coupling constants can be eliminated, just as for thermalizing thermometry. We note that some knowledge of the hot bath temperature is required for our scheme. However, the cold temperature can be determined with high precision even if the hot temperature measurement is noisy. Our scheme can thus be seen as a method to turn an uncertain measurement of warm temperatures into precise measurements of cold temperatures {working similarly to a Wheatstone bridge, where resistors of known resistances are used to determine an unknown resistance}.

The rest of this letter is structured as follows. After describing the working principle of the proposed thermometer in more detail, we discuss an implementation in a circuit QED architecture, which allows for precise thermometry {of a microwave resonator} down to $\sim 15$\,mK using realistic parameters.
Finally, we investigate the precision of the thermometer using the quantum Fisher information.

\begin{figure}[b!]
  \centering
  \includegraphics[width=\columnwidth]{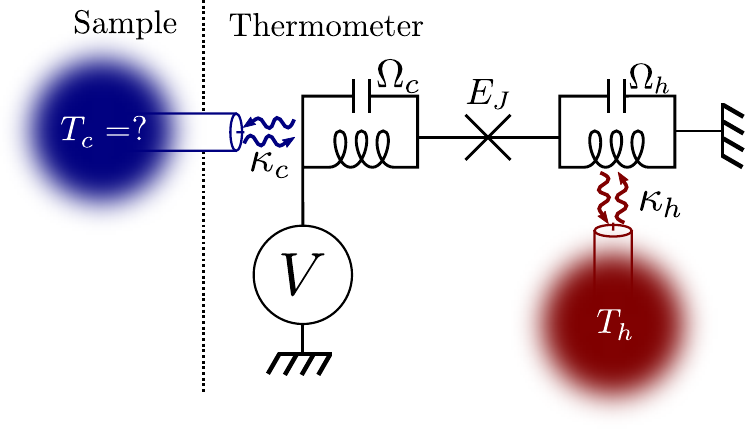}
  \caption{Circuit QED implementation of the thermometer. Two harmonic oscillators with frequencies $\Omega_h$ and $\Omega_c$ are coupled to thermal baths at temperatures $T_c$ and $T_h$ respectively and to each other through a Josephson junction. The external bias voltage $V$ ensures that the cold bath is being cooled while determining $T_c$.}
  \label{fig:cqed}
\end{figure}
\begin{table}[b!]
\centering
\def\arraystretch{1.3}
\begin{tabular}{|c|c|c|c|c|c|c|c|c|c| }
\hline
  \,$\Omega_h/2\pi$\, & \,$\Omega_c/2\pi$\, & \,$\kappa/2\pi$\, &\, $E_J/2\pi$\, &
 \, $\lambda$\, &\, $\Delta I$\,&\, $\Delta T_h$\,\\\hline\hline
     \,$8.5$\,GHZ\, &\, $1$\,GHZ\, &\, $0.06$\,GHz\, & \,$0.2\,$GHz \,&
      \,$0.3$\, &\,$0.3\,$pA \,&\,$10\,$mK \,\\
\hline
\end{tabular}
\caption{Realistic Parameters for operating the proposed thermometer. Here $\kappa=\kappa_h=\kappa_c$ and $\lambda=\lambda_h=\lambda_c$.}
\label{tab:params}
\end{table}

\begin{figure*}[t!]
  \centering
  \includegraphics[width=0.9\textwidth]{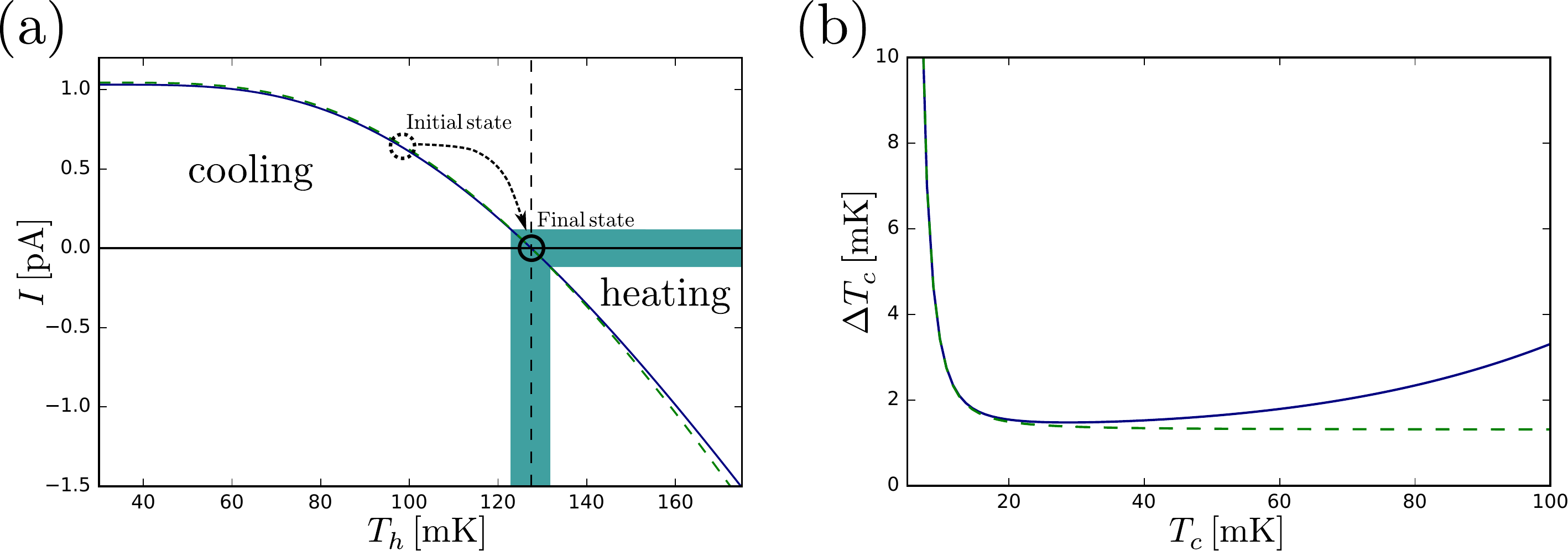}
  \caption{Performance of a circuit QED implementation. (a) Charge current as a function of the hot temperature $T_h$ for fixed $T_c=15\,$mK. The refrigerator is initiated at $T_h$ below the Carnot point (dotted circle) leading to cooling of the cold bath. $T_h$ is then increased until the current vanishes (solid circle). This point can only be determined up to a certain error $\Delta I$ which induces an uncertainty in the final $T_h$ (shaded area) reducing the accuracy of the measurement of $T_c$. The bigger $|\partial_{T_h} I|$, the smaller the induced error. (b) Error in the temperature estimation. Two sources of error limit the accuracy of measuring $T_c$. The Carnot point can only approximately be reached due to a finite resolution in the current measurement $\Delta I$ [see panel (a)] and the hot temperature can only be measured with an accuracy $\Delta T_h$. The total resulting error $\Delta T_c$ is given in Eq.~\eqref{eq:errortc} and plotted as a function of $T_c$. Measurement errors of $\Delta I = 0.3\,$pA and $\Delta T_h=10\,$mK lead to $\Delta T_c < 2\,$mK down to temperatures of $15\,$mK. Blue (solid) lines are numeric solutions, green (dashed) lines are obtained analytically using a simplified model \cite{supplementarx}.}
  \label{fig:implementation}
\end{figure*}

\textit{Scheme.---} We now turn to a more detailed description of our thermometer. 
The thermal bias and the power source induce energy flows denoted by $J_c$ (heat flow into the cold bath), $J_h$ (heat flow into the hot bath), and $P$ (power consumption, see Fig.~\ref{fig:sketch}). The power consumed by the machine will in general be a function of the temperatures as well as the model-dependent parameters. Inverting this relationship, the cold temperature {can be written as}
\begin{equation}
\label{eq:tcf}
T_c = f(P,T_h) .
\end{equation}
Here, $f$ might have a complicated dependence on the coupling constants or any other model dependent parameters. However, the relation simplifies for machines that perform at the Otto efficiency and exhibit a Carnot point \cite{mahler:book}. When the machine is operated as a heat engine, the efficiency is defined as $\eta = P/J_h$. The Otto efficiency is given by $\eta=1-\Omega_c/\Omega_h$
where $\Omega_c$, $\Omega_h$ are frequencies that depend on the architecture of the machine (see Fig.~\ref{fig:cqed}). For thermal machines that exhibit a Carnot point, setting the frequencies and temperatures such that
\begin{equation}
\label{eq:otto_carnot}
\eta = 1-\frac{\Omega_c}{\Omega_h} = 1-\frac{T_c}{T_h} = \eta_C,
\end{equation}
results in vanishing energy flows. Examples of thermal machines which operate at the Otto efficiency and exhibit a Carnot point are discussed in Refs.~\cite{kosloff:1984,feldmann:2003,quan:2007,niskanen:2007,henrich:2007,hofer:2016prb,campisi:2016}. We note that whenever the frequencies and temperatures are such that $\eta>\eta_C$, the machine operates as a refrigerator.
At the Carnot point, we find the simple relation
\begin{equation}
\label{eq:relationcarnot}
T_c=\frac{\Omega_c}{\Omega_h} T_h ,
\end{equation}
implying that $f(P=0,T_h)=T_h \Omega_c/\Omega_h$ is independent of any coupling constants.
{Thomson proposed to use the above relation to determine an absolute temperature scale \cite{kelvin:1848}. Here we use the same relation as a key ingredient in low temperature thermometry (see supplemental material for a discussion on imperfections that prevent Eq.~\eqref{eq:relationcarnot} \cite{supplementarx}). In order to reach the Carnot point, one can either modify the frequencies $\Omega_\alpha$ or the temperature $T_h$ associated with the thermometer. Here we focus on the case where the frequencies remain fixed but one has some control over $T_h>T_c$.}
The cold temperature can then be determined using the following strategy (see Fig.~\ref{fig:implementation})
\begin{enumerate}
\item Initiate the machine to act as a refrigerator, i.e. $T_h<T_c\Omega_h/\Omega_c$, and monitor $P$
\item Increase $T_h$ until $P=0$ is reached
\item Measure $T_h$
\item Determine $T_c$ using Eq.~\eqref{eq:relationcarnot}
\end{enumerate}
{In this scheme,} two quantities are measured to determine $T_c$: The power consumption $P$ and the hot temperature $T_h$. Both of these measurements are accompanied by errors, $\Delta P$ and $\Delta T_h$. Simple error propagation yields (assuming independent errors)
\begin{equation}
\label{eq:errortc}
\Delta T_c=\sqrt{\left(\frac{\partial f}{\partial P}\right)^2(\Delta P)^2+\left(\frac{\partial f}{\partial T_h}\right)^2(\Delta T_h)^2}.
\end{equation}
The error thus depends on the derivatives of $f$ which generally depend on model-specific parameters. Note however that $\partial_{T_h} f|_{P=0}=\Omega_c/\Omega_h$ implying that the error induced by the measurement of $T_h$ only depends on the frequencies. Any uncertainty in the measurement of $T_h$ can thus be compensated by increasing the ratio $\Omega_h/\Omega_c$ and thus does not represent a fundamental limit.

\textit{Circuit QED implementation.---} We now turn to an implementation of these ideas, considering the heat engine proposed in Ref.~\cite{hofer:2016prb} and sketched in Fig.~\ref{fig:cqed}. In this machine, the quantum system consists of two $LC$-oscillators {with frequencies $\Omega_c$ and $\Omega_h$} coupled to each other through a Josephson junction. 
{Such a system has recently been implemented experimentally, investigating the emission of non-classical radiation \cite{westig:2017}. See also Refs.~\cite{holst:1994,basset:2010,hofheinz:2011} for related experiments.}
Each oscillator is coupled individually to a heat bath, {one of which is the sample at temperature $T_c$,} and the power source is provided by an external voltage bias $V$. We note that a similar setup (without any temperature bias however) has been considered for thermometry in Ref.~\cite{saira:2016}. The Hamiltonian describing the system reads (in a rotating frame)
\begin{equation}
\label{eq:ham}
\hat{H}= \frac{E_J}{2}\bigg(\hat{a}_h^\dag\hat{A}_h\hat{A}_c\hat{a}_c+H.c.\bigg),
\end{equation}
where $E_J$ is the Josephson energy, {$\hat{a}_\alpha$ annihilates a photon in the oscillator with frequency $\Omega_\alpha$} and the non-linear operators $\hat{A}_\alpha$ are defined as
\begin{equation}
\label{eq:aops}
\hat{A}_\alpha=2\lambda_\alpha e^{-2\lambda_\alpha^2}\sum\limits_{n_\alpha=0}^{\infty}\frac{L_{n_\alpha}^{(1)}(4\lambda_\alpha^2)}{n_\alpha+1}|n_\alpha\rangle\langle n_\alpha |,
\end{equation}
with $L_n^{(k)}(x)$ denoting the generalized Laguerre polynomials and {where we defined the Fock states $\hat{a}^\dagger_\alpha\hat{a}_\alpha|n_\alpha\rangle=n_\alpha|n_\alpha\rangle$}. We note that Eq.~\eqref{eq:ham} is derived using a rotating wave approximation which holds under the resonance condition {(for details, see Ref.~\cite{hofer:2016prb})}
\begin{equation}
\label{eq:resonance}
2eV=\Omega_h-\Omega_c.
\end{equation}
The evolution of the system in contact with the thermal baths is captured by a local Lindblad master equation
\begin{equation}
\label{eq:master}
\begin{aligned}
\partial_t\hat{\rho}=-i[\hat{H},\hat{\rho}]&+\kappa_h(n_B^h+1)\mathcal{D}[\hat{a}_h]\hat{\rho}+\kappa_hn_B^h\mathcal{D}[\hat{a}_h^\dag]\hat{\rho}\\&+\kappa_c(n_B^c+1)\mathcal{D}[\hat{a}_c]\hat{\rho}+\kappa_hn_B^c\mathcal{D}[\hat{a}_c^\dag]\hat{\rho},
\end{aligned}
\end{equation}
where we defined $\mathcal{D}[\hat{A}]\hat{\rho}=\hat{A}\hat{\rho}\hat{A}^\dag-\{\hat{A}^\dag\hat{A},\hat{\rho}\}/2$, $\kappa_\alpha$ denotes the energy damping rate associated with the bath $\alpha$, and $n_B^\alpha=[\exp({\frac{\Omega_\alpha}{k_BT_\alpha}})-1]^{-1}$ the corresponding occupation number.
{In accordance with existing theory \cite{gramich:2013} and experiment \cite{hofheinz:2011}, we neglect voltage fluctuations arising from a low-frequency environment.} {We note that the local master equation, Eq.~\eqref{eq:master}, was recently shown to capture the thermodynamics of the considered heat engine very well \cite{hofer:2017,gonzalez:2017}. Alternatively one may also consider a master equation based on a Floquet formalism, see Refs.~\cite{alicki:2006,levy:2012pre,szczygielski:2013}}.

The power consumption of the machine is $P=IV$, where $I=\langle \hat{I} \rangle$ is the (dc) electrical current with the current operator
\begin{equation}
\label{eq:curr}
\hat{I}=-\frac{I_c}{2i}\bigg(\hat{a}_c^\dag\hat{A}_c\hat{A}_h\hat{a}_h-H.c.\bigg),
\end{equation}
and $I_c=2eE_J$ the critical current. The mean heat currents are defined as
\begin{equation}
\label{eq:heatcurr}
J_\alpha=\Omega_\alpha\kappa_\alpha\left(\langle \hat{n}_\alpha\rangle-n_B^\alpha\right).
\end{equation}
{All averages are taken with respect to the steady-state solution of Eq.~\eqref{eq:master}. For a more detailed discussion on the working principle of the heat engine and the involved approximations, we refer the reader to Ref.~\cite{hofer:2016prb}.} It can be shown that this machine does perform at the Otto efficiency and can reach the Carnot efficiency at vanishing power. Furthermore, a tunable hot temperature could be implemented by heating one of the $LC$-oscillators using a microwave antenna. {Therefore, this circuit QED engine exhibits all the features required to perform thermometry as discussed above.}

\figref{fig:implementation}\,(a) shows the electrical current as a function of the hot temperature and sketches the scheme for measuring $T_c$. The error of the measurement is plotted in \figref{fig:implementation}\,(b), where the parameters (including the uncertainties) are given in \tabref{tab:params}. We find a precision of $\Delta T_c\lesssim 2\,$mK for temperatures down to $T_c\sim15\,$mK. In accordance with Ref.~\cite{hofer:2016prb}, we find good quantitative agreement with an approximate model obtained from Eq.~\eqref{eq:ham} by replacing the non-linear operators $\hat{A}_\alpha$ by constants times the identity \cite{supplementarx}. This model can be solved analytically and is used below to estimate the performance of our scheme using the quantum Fisher information.
{We note that instead of the electrical current, one could also measure the heat currents to determine the Carnot point.}

{Throughout this paper, we consider the thermometer as a device to determine the temperature of the cold bath. However, in this particular implementation, the thermometer measures the temperature of the microwave mode with frequency $\Omega_c$ (see supplemental material for a discussion where this temperature is not equal to the bath temperature \cite{supplementarx}). In situations where thermalization between different components of an experimental setup is difficult to achieve, our proposal thus provides a promising route to determine the physically relevant temperature. Another possibility for measuring the temperature in the circuit is to perform shot noise thermometry, where a magnetic field needs to be applied to bring the junction to the normal state, possibly influencing the temperature. Such a measurement was performed in Ref.~\cite{hofheinz:2011} to determine the temperature of a microwave resonator coupled to a Josephson junction. In general, the precision obtained in our proposal compares well with electronic out-of-equilibrium thermometry \cite{iftikhar:2016}.}

\begin{figure}[t!]
  \centering
  \includegraphics[width=0.95\linewidth]{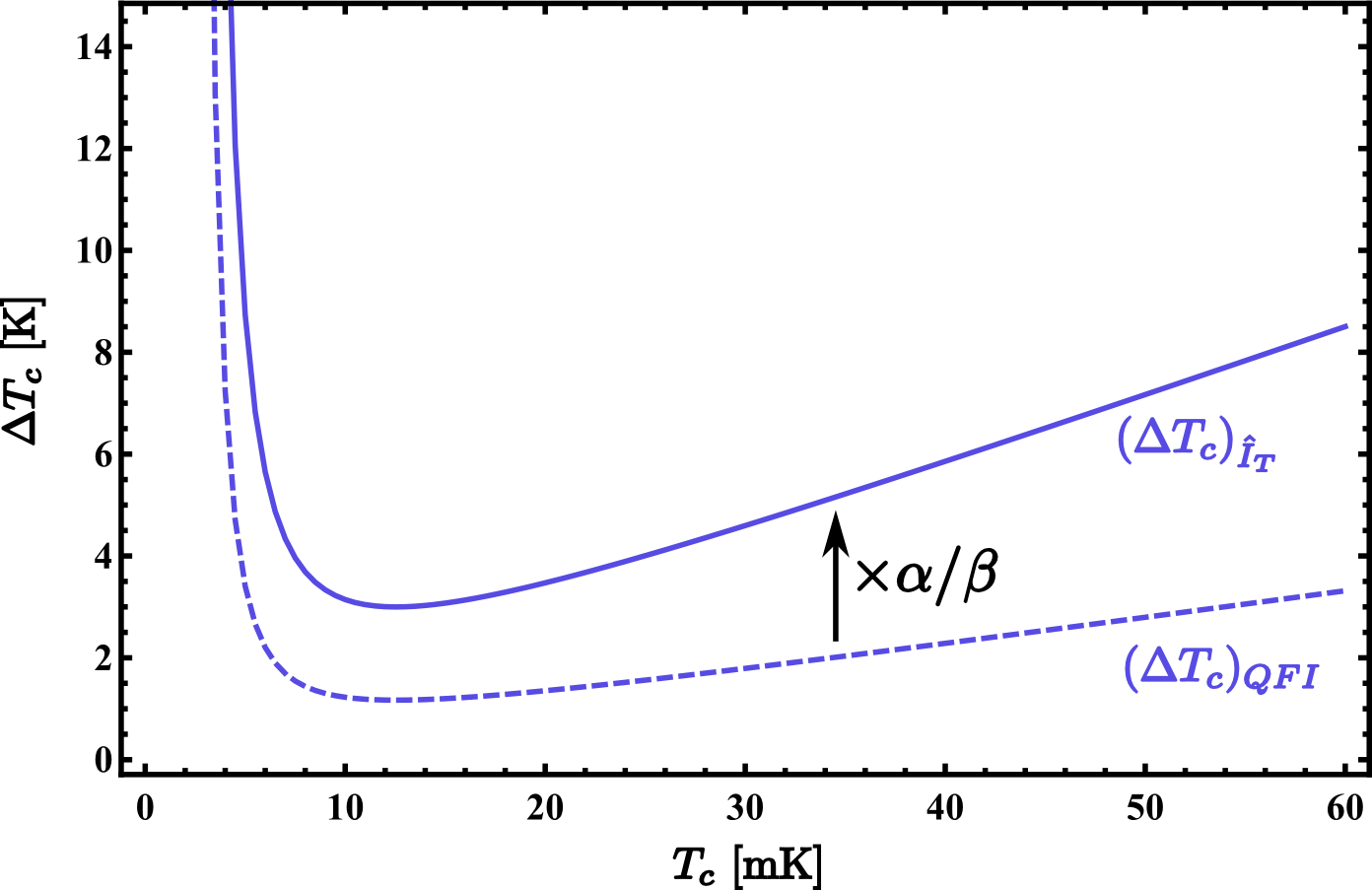}
  \caption{Single-shot precision from a current measurement (blue solid) and from the QFI (blue dashed) at the Carnot point, for the simple model with oscillator frequencies and couplings given in \tabref{tab:params}, and fitting parameter $g=E_J/8$.}
  \label{fig:qfi}
\end{figure}

\textit{Quantum Fisher information.---} As already mentioned, the measurement error resulting from the hot temperature measurement is not of a fundamental nature since we can in principle reduce it by increasing $\Omega_h/\Omega_c$ (note however that this implies an increase of $T_h$ at the Carnot point). In order to understand how well the current measurement is doing in terms of temperature estimation, we turn to the quantum Fisher information.

The steady-state solution of Eq.~\eqref{eq:master} defines a family of states as a function of $T_c$ (as well as all the other parameters in the setup). The quantum Fisher information (QFI) w.r.t~$T_c$ is a measure of the sensitivity of the state to changes in this parameter \cite{holevo:1982}. Through the quantum Cramer-Rao bound, it provides a lower bound on the mean-squared error in estimating $T_c$ from any possible measurement \cite{braunstein:1994}. Specifically,
\begin{equation}
\label{eq:cramerrao}
(\Delta T_c)^2 \geq \frac{1}{\nu F_{T_c}} \equiv \frac{1}{\nu} (\Delta T_c)^2_{QFI},
\end{equation}
where $F_{T_c}$ is the QFI w.r.t.~$T_c$ and $\nu$ is the number of independent repetitions of the experiment. The bound is relevant (and can be saturated) in the local estimation regime, where the prior on the estimated parameter is narrow and many repetitions are performed. In the same regime, for measuring a specific observable $\hat O$, the attainable precision is given by the error propagation formula
{
\begin{equation}
\label{eq:errorprop}
\frac{1}{\nu}(\Delta T_c)^2_{\hat O} \equiv \frac{(\Delta \hat O)_\rho^2}{\nu|\partial_{T_c} \langle \hat O \rangle_\rho|^2} .
\end{equation}}
Hence, by substituting the current operator for $\hat O$ and comparing $(\Delta T_c)_{QFI}$ and $(\Delta T_c)_{\hat I}$ evaluated in the steady state at the Carnot point, we can get an idea of how close the current measurement is to being optimal. Note that this neglects any errors in other parameters as well as the fact that our strategy is based on continuous measurements rather than projective measurements considered in Eq.~\eqref{eq:errorprop}. Although we should thus not expect this calculation to give us the actual uncertainty obtained in an experiment, it does tell us whether the current is a good choice of observable for estimating $T_c$.

To perform the comparison, we first need to find the steady-state solution of Eq.~\eqref{eq:master}. This can be done analytically for the approximate model discussed in the supplemental material \cite{supplementarx} yielding
\begin{equation}
\label{eq.Iprec}
(\Delta T_c)_{{\hat I}_T}  = \alpha \frac{T_c^2}{\Omega_c} \sinh\left(\frac{\Omega_c}{2T_c}\right) ,
\end{equation}
and
\begin{equation}
\label{eq.QFIprec}
(\Delta T_c)_{QFI} = \beta \frac{T_c^2}{\Omega_c} \sinh\left(\frac{\Omega_c}{2T_c}\right),
\end{equation}
where the parameters $\alpha$ and $\beta$ depend on the coupling constants and are given in the supplemental material \cite{supplementarx}.
We see that the current-measurement precision and the optimal precision from the QFI have exactly the same functional behavior with temperature and energy, but depend differently on the coupling constants. For any choice of couplings, $\alpha \geq \beta$ such that $(\Delta T_c)_{{\hat I}_T} \geq (\Delta T_c)_{QFI}$. Both expressions are plotted in \figref{fig:qfi}. 
{The reason that the current operator is a good observable for determining $T_c$ is ultimately due to the fact that the current is very sensitive to the occupation number in the oscillators. Although the occupation numbers depend only very weakly on temperature at sufficiently low temperatures, measuring occupation numbers is in many scenarios still the optimal choice \cite{campbell:2017}.}

The smallest possible value of $\alpha$ is $4\sqrt{2}$, attained for $\kappa_c = \kappa_h = 2g$ (where $g$ is the interaction strength in the approximate model). For the parameters in \tabref{tab:params}, $\alpha/4\sqrt{2} \approx 1.02$, and so the current measurement is close to its best performance in this regime. Furthermore, $\alpha/\beta \approx 2.55$, hence it is also not far from the optimal precision obtainable by any possible measurement. The smallest possible value of $\alpha/\beta$ is $\sqrt{2+\sqrt{2}} \approx 1.85$, however it is attained in the weak coupling limit $\kappa_c \rightarrow 0$, where no information about $T_c$ can be extracted and both $\alpha$ and $\beta$ diverge. In the opposite limit where the hot reservoir is disconnected, $\kappa_h \rightarrow 0$, we have $\alpha \rightarrow \infty$ while $\beta \rightarrow \sqrt{2}$ reaches its minimal value. This reflects the fact that the current vanishes if the system is coupled to a single bath only, while even if there is no coupling to the hot bath, the steady state of the system still contains information about $T_c$.

\textit{Conclusion.---} In conclusion, we investigated the use of thermal machines as thermometers. The non-equilibrium nature of the steady state in these systems allows for simultaneously cooling the cold bath while determining its temperature. Furthermore, our scheme only requires the measurement of the power consumption and the hot temperature, both of which do not require any projective measurements on the quantum system which are difficult to implement. For realistic parameters, an implementation in circuit QED allows for precise thermometry ($\Delta T_c\lesssim 2$\,mK) {of a microwave resonator mode} down to small temperatures ($T_c\sim 15\,$mK). While lower values have been reported (see e.g. Ref.~\cite{iftikhar:2016} where precise thermometry down to $6$\,mK has been reported) our scheme is of particular interest when {thermalization between different components of an experimental setup cannot be guaranteed and} heating of the sample has to be avoided. Our proposal can be readily adapted to other architectures.

\textit{Acknowledgements.---} We acknowledge interesting discussions with L. Correa. P. H., J. B., and N. B. acknowledge the Swiss National
Science  Foundation  (starting  Grant  DIAQ,  Grant No. 200021\_169002, and QSIT). M. P.-L. acknowledges the Alexander von Humboldt Foundation, the CELLEX-ICFO-MPQ research fellowships, the Spanish MINECO (Grants No. QIBEQI FIS2016-80773-P and No. SEV-2015-0522), and the Generalitat de Catalunya (Grants No. SGR875 and CERCA Programme).

\bibliography{biblio}

\begin{thebibliography}{67}%
\makeatletter
\providecommand \@ifxundefined [1]{%
 \@ifx{#1\undefined}
}%
\providecommand \@ifnum [1]{%
 \ifnum #1\expandafter \@firstoftwo
 \else \expandafter \@secondoftwo
 \fi
}%
\providecommand \@ifx [1]{%
 \ifx #1\expandafter \@firstoftwo
 \else \expandafter \@secondoftwo
 \fi
}%
\providecommand \natexlab [1]{#1}%
\providecommand \enquote  [1]{``#1''}%
\providecommand \bibnamefont  [1]{#1}%
\providecommand \bibfnamefont [1]{#1}%
\providecommand \citenamefont [1]{#1}%
\providecommand \href@noop [0]{\@secondoftwo}%
\providecommand \href [0]{\begingroup \@sanitize@url \@href}%
\providecommand \@href[1]{\@@startlink{#1}\@@href}%
\providecommand \@@href[1]{\endgroup#1\@@endlink}%
\providecommand \@sanitize@url [0]{\catcode `\\12\catcode `\$12\catcode
  `\&12\catcode `\#12\catcode `\^12\catcode `\_12\catcode `\%12\relax}%
\providecommand \@@startlink[1]{}%
\providecommand \@@endlink[0]{}%
\providecommand \url  [0]{\begingroup\@sanitize@url \@url }%
\providecommand \@url [1]{\endgroup\@href {#1}{\urlprefix }}%
\providecommand \urlprefix  [0]{URL }%
\providecommand \Eprint [0]{\href }%
\providecommand \doibase [0]{http://dx.doi.org/}%
\providecommand \selectlanguage [0]{\@gobble}%
\providecommand \bibinfo  [0]{\@secondoftwo}%
\providecommand \bibfield  [0]{\@secondoftwo}%
\providecommand \translation [1]{[#1]}%
\providecommand \BibitemOpen [0]{}%
\providecommand \bibitemStop [0]{}%
\providecommand \bibitemNoStop [0]{.\EOS\space}%
\providecommand \EOS [0]{\spacefactor3000\relax}%
\providecommand \BibitemShut  [1]{\csname bibitem#1\endcsname}%
\let\auto@bib@innerbib\@empty
\bibitem [{\citenamefont {Carlos}\ and\ \citenamefont
  {Palacio}(2016)}]{carlospalacio:2016}%
  \BibitemOpen
  \bibinfo {editor} {\bibfnamefont {L.~D.}\ \bibnamefont {Carlos}}\ and\
  \bibinfo {editor} {\bibfnamefont {F.}~\bibnamefont {Palacio}},\ eds.,\ \href
  {\doibase 10.1039/9781782622031} {\emph {\bibinfo {title} {Thermometry at the
  Nanoscale}}}\ (\bibinfo  {publisher} {The Royal Society of Chemistry},\
  \bibinfo {year} {2016})\BibitemShut {NoStop}%
\bibitem [{\citenamefont {Walker}\ \emph {et~al.}(2003)\citenamefont {Walker},
  \citenamefont {Sundar}, \citenamefont {Rudzinski}, \citenamefont {Wun},
  \citenamefont {Bawendi},\ and\ \citenamefont {Nocera}}]{walker:2003}%
  \BibitemOpen
  \bibfield  {author} {\bibinfo {author} {\bibfnamefont {G.~W.}\ \bibnamefont
  {Walker}}, \bibinfo {author} {\bibfnamefont {V.~C.}\ \bibnamefont {Sundar}},
  \bibinfo {author} {\bibfnamefont {C.~M.}\ \bibnamefont {Rudzinski}}, \bibinfo
  {author} {\bibfnamefont {A~W.}\ \bibnamefont {Wun}}, \bibinfo {author}
  {\bibfnamefont {M~G.}\ \bibnamefont {Bawendi}}, \ and\ \bibinfo {author}
  {\bibfnamefont {D~G.}\ \bibnamefont {Nocera}},\ }\bibfield  {title} {\enquote
  {\bibinfo {title} {Quantum-dot optical temperature probes},}\ }\href
  {\doibase 10.1063/1.1620686} {\bibfield  {journal} {\bibinfo  {journal}
  {Appl. Phys. Lett.}\ }\textbf {\bibinfo {volume} {83}},\ \bibinfo {pages}
  {3555} (\bibinfo {year} {2003})}\BibitemShut {NoStop}%
\bibitem [{\citenamefont {Seilmeier}\ \emph {et~al.}(2014)\citenamefont
  {Seilmeier}, \citenamefont {Hauck}, \citenamefont {Schubert}, \citenamefont
  {Schinner}, \citenamefont {Beavan},\ and\ \citenamefont
  {H\"ogele}}]{seilmeier:2014}%
  \BibitemOpen
  \bibfield  {author} {\bibinfo {author} {\bibfnamefont {F.}~\bibnamefont
  {Seilmeier}}, \bibinfo {author} {\bibfnamefont {M.}~\bibnamefont {Hauck}},
  \bibinfo {author} {\bibfnamefont {E.}~\bibnamefont {Schubert}}, \bibinfo
  {author} {\bibfnamefont {G.~J.}\ \bibnamefont {Schinner}}, \bibinfo {author}
  {\bibfnamefont {S.~E.}\ \bibnamefont {Beavan}}, \ and\ \bibinfo {author}
  {\bibfnamefont {A.}~\bibnamefont {H\"ogele}},\ }\bibfield  {title} {\enquote
  {\bibinfo {title} {Optical thermometry of an electron reservoir coupled to a
  single quantum dot in the millikelvin range},}\ }\href {\doibase
  10.1103/PhysRevApplied.2.024002} {\bibfield  {journal} {\bibinfo  {journal}
  {Phys. Rev. Applied}\ }\textbf {\bibinfo {volume} {2}},\ \bibinfo {pages}
  {024002} (\bibinfo {year} {2014})}\BibitemShut {NoStop}%
\bibitem [{\citenamefont {Haupt}\ \emph {et~al.}(2014)\citenamefont {Haupt},
  \citenamefont {Imamoglu},\ and\ \citenamefont {Kroner}}]{haupt:2014}%
  \BibitemOpen
  \bibfield  {author} {\bibinfo {author} {\bibfnamefont {F.}~\bibnamefont
  {Haupt}}, \bibinfo {author} {\bibfnamefont {A.}~\bibnamefont {Imamoglu}}, \
  and\ \bibinfo {author} {\bibfnamefont {M.}~\bibnamefont {Kroner}},\
  }\bibfield  {title} {\enquote {\bibinfo {title} {Single quantum dot as an
  optical thermometer for millikelvin temperatures},}\ }\href {\doibase
  10.1103/PhysRevApplied.2.024001} {\bibfield  {journal} {\bibinfo  {journal}
  {Phys. Rev. Applied}\ }\textbf {\bibinfo {volume} {2}},\ \bibinfo {pages}
  {024001} (\bibinfo {year} {2014})}\BibitemShut {NoStop}%
\bibitem [{\citenamefont {Toyli}\ \emph {et~al.}(2013)\citenamefont {Toyli},
  \citenamefont {de~las Casas}, \citenamefont {Christle}, \citenamefont
  {Dobrovitski},\ and\ \citenamefont {Awschalom}}]{toyli:2013}%
  \BibitemOpen
  \bibfield  {author} {\bibinfo {author} {\bibfnamefont {D.~M.}\ \bibnamefont
  {Toyli}}, \bibinfo {author} {\bibfnamefont {C.~F.}\ \bibnamefont {de~las
  Casas}}, \bibinfo {author} {\bibfnamefont {D.~J.}\ \bibnamefont {Christle}},
  \bibinfo {author} {\bibfnamefont {V.~V.}\ \bibnamefont {Dobrovitski}}, \ and\
  \bibinfo {author} {\bibfnamefont {D.~D.}\ \bibnamefont {Awschalom}},\
  }\bibfield  {title} {\enquote {\bibinfo {title} {Fluorescence thermometry
  enhanced by the quantum coherence of single spins in diamond},}\ }\href
  {\doibase 10.1073/pnas.1306825110} {\bibfield  {journal} {\bibinfo  {journal}
  {Proc. Natl. Acad. Sci. U.S.A.}\ }\textbf {\bibinfo {volume} {110}},\
  \bibinfo {pages} {8417} (\bibinfo {year} {2013})}\BibitemShut {NoStop}%
\bibitem [{\citenamefont {Kucsko}\ \emph {et~al.}(2013)\citenamefont {Kucsko},
  \citenamefont {Maurer}, \citenamefont {Yao}, \citenamefont {Kubo},
  \citenamefont {Noh}, \citenamefont {Lo}, \citenamefont {Park},\ and\
  \citenamefont {Lukin}}]{kucsko:2013}%
  \BibitemOpen
  \bibfield  {author} {\bibinfo {author} {\bibfnamefont {G.}~\bibnamefont
  {Kucsko}}, \bibinfo {author} {\bibfnamefont {P.~C.}\ \bibnamefont {Maurer}},
  \bibinfo {author} {\bibfnamefont {N.~Y.}\ \bibnamefont {Yao}}, \bibinfo
  {author} {\bibfnamefont {M.}~\bibnamefont {Kubo}}, \bibinfo {author}
  {\bibfnamefont {H.~J.}\ \bibnamefont {Noh}}, \bibinfo {author} {\bibfnamefont
  {P.~K.}\ \bibnamefont {Lo}}, \bibinfo {author} {\bibfnamefont
  {H.}~\bibnamefont {Park}}, \ and\ \bibinfo {author} {\bibfnamefont {M.~D.}\
  \bibnamefont {Lukin}},\ }\bibfield  {title} {\enquote {\bibinfo {title}
  {Nanometre-scale thermometry in a living cell},}\ }\href
  {http://dx.doi.org/10.1038/nature12373} {\bibfield  {journal} {\bibinfo
  {journal} {Nature}\ }\textbf {\bibinfo {volume} {500}},\ \bibinfo {pages}
  {54} (\bibinfo {year} {2013})}\BibitemShut {NoStop}%
\bibitem [{\citenamefont {Neumann}\ \emph {et~al.}(2013)\citenamefont
  {Neumann}, \citenamefont {Jakobi}, \citenamefont {Dolde}, \citenamefont
  {Burk}, \citenamefont {Reuter}, \citenamefont {Waldherr}, \citenamefont
  {Honert}, \citenamefont {Wolf}, \citenamefont {Brunner}, \citenamefont
  {Shim}, \citenamefont {Suter}, \citenamefont {Sumiya}, \citenamefont
  {Isoya},\ and\ \citenamefont {Wrachtrup}}]{neumann:2013}%
  \BibitemOpen
  \bibfield  {author} {\bibinfo {author} {\bibfnamefont {P.}~\bibnamefont
  {Neumann}}, \bibinfo {author} {\bibfnamefont {I.}~\bibnamefont {Jakobi}},
  \bibinfo {author} {\bibfnamefont {F.}~\bibnamefont {Dolde}}, \bibinfo
  {author} {\bibfnamefont {C.}~\bibnamefont {Burk}}, \bibinfo {author}
  {\bibfnamefont {R.}~\bibnamefont {Reuter}}, \bibinfo {author} {\bibfnamefont
  {G.}~\bibnamefont {Waldherr}}, \bibinfo {author} {\bibfnamefont
  {J.}~\bibnamefont {Honert}}, \bibinfo {author} {\bibfnamefont
  {T.}~\bibnamefont {Wolf}}, \bibinfo {author} {\bibfnamefont {A.}~\bibnamefont
  {Brunner}}, \bibinfo {author} {\bibfnamefont {J.~H.}\ \bibnamefont {Shim}},
  \bibinfo {author} {\bibfnamefont {D.}~\bibnamefont {Suter}}, \bibinfo
  {author} {\bibfnamefont {H.}~\bibnamefont {Sumiya}}, \bibinfo {author}
  {\bibfnamefont {J.}~\bibnamefont {Isoya}}, \ and\ \bibinfo {author}
  {\bibfnamefont {J.}~\bibnamefont {Wrachtrup}},\ }\bibfield  {title} {\enquote
  {\bibinfo {title} {High-precision nanoscale temperature sensing using single
  defects in diamond},}\ }\href {\doibase 10.1021/nl401216y} {\bibfield
  {journal} {\bibinfo  {journal} {Nano Lett.}\ }\textbf {\bibinfo {volume}
  {13}},\ \bibinfo {pages} {2738} (\bibinfo {year} {2013})}\BibitemShut
  {NoStop}%
\bibitem [{\citenamefont {Halbertal}\ \emph {et~al.}(2016)\citenamefont
  {Halbertal}, \citenamefont {Cuppens}, \citenamefont {Shalom}, \citenamefont
  {Embon}, \citenamefont {Shadmi}, \citenamefont {Anahory}, \citenamefont
  {Naren}, \citenamefont {Sarkar}, \citenamefont {Uri}, \citenamefont {Ronen},
  \citenamefont {Myasoedov}, \citenamefont {Levitov}, \citenamefont
  {Joselevich}, \citenamefont {Geim},\ and\ \citenamefont
  {Zeldov}}]{halbertal:2016}%
  \BibitemOpen
  \bibfield  {author} {\bibinfo {author} {\bibfnamefont {D.}~\bibnamefont
  {Halbertal}}, \bibinfo {author} {\bibfnamefont {J.}~\bibnamefont {Cuppens}},
  \bibinfo {author} {\bibfnamefont {M.~Ben}\ \bibnamefont {Shalom}}, \bibinfo
  {author} {\bibfnamefont {L.}~\bibnamefont {Embon}}, \bibinfo {author}
  {\bibfnamefont {N.}~\bibnamefont {Shadmi}}, \bibinfo {author} {\bibfnamefont
  {Y.}~\bibnamefont {Anahory}}, \bibinfo {author} {\bibfnamefont {H.~R.}\
  \bibnamefont {Naren}}, \bibinfo {author} {\bibfnamefont {J.}~\bibnamefont
  {Sarkar}}, \bibinfo {author} {\bibfnamefont {A.}~\bibnamefont {Uri}},
  \bibinfo {author} {\bibfnamefont {Y.}~\bibnamefont {Ronen}}, \bibinfo
  {author} {\bibfnamefont {Y.}~\bibnamefont {Myasoedov}}, \bibinfo {author}
  {\bibfnamefont {L.~S.}\ \bibnamefont {Levitov}}, \bibinfo {author}
  {\bibfnamefont {E.}~\bibnamefont {Joselevich}}, \bibinfo {author}
  {\bibfnamefont {A.~K.}\ \bibnamefont {Geim}}, \ and\ \bibinfo {author}
  {\bibfnamefont {E.}~\bibnamefont {Zeldov}},\ }\bibfield  {title} {\enquote
  {\bibinfo {title} {Nanoscale thermal imaging of dissipation in quantum
  systems},}\ }\href {http://dx.doi.org/10.1038/nature19843} {\bibfield
  {journal} {\bibinfo  {journal} {Nature}\ }\textbf {\bibinfo {volume} {539}},\
  \bibinfo {pages} {407} (\bibinfo {year} {2016})}\BibitemShut {NoStop}%
\bibitem [{\citenamefont {Donner}\ \emph {et~al.}(2012)\citenamefont {Donner},
  \citenamefont {Thompson}, \citenamefont {Kreuzer}, \citenamefont {Baffou},\
  and\ \citenamefont {Quidant}}]{donner:2012}%
  \BibitemOpen
  \bibfield  {author} {\bibinfo {author} {\bibfnamefont {J.~S.}\ \bibnamefont
  {Donner}}, \bibinfo {author} {\bibfnamefont {S.~A.}\ \bibnamefont
  {Thompson}}, \bibinfo {author} {\bibfnamefont {M.~P.}\ \bibnamefont
  {Kreuzer}}, \bibinfo {author} {\bibfnamefont {G.}~\bibnamefont {Baffou}}, \
  and\ \bibinfo {author} {\bibfnamefont {R.}~\bibnamefont {Quidant}},\
  }\bibfield  {title} {\enquote {\bibinfo {title} {Mapping intracellular
  temperature using green fluorescent protein},}\ }\href {\doibase
  10.1021/nl300389y} {\bibfield  {journal} {\bibinfo  {journal} {Nano Lett.}\
  }\textbf {\bibinfo {volume} {12}},\ \bibinfo {pages} {2107} (\bibinfo {year}
  {2012})}\BibitemShut {NoStop}%
\bibitem [{\citenamefont {Goold}\ \emph {et~al.}(2016)\citenamefont {Goold},
  \citenamefont {Huber}, \citenamefont {Riera}, \citenamefont {del Rio},\ and\
  \citenamefont {Skrzypczyk}}]{goold:2016}%
  \BibitemOpen
  \bibfield  {author} {\bibinfo {author} {\bibfnamefont {J.}~\bibnamefont
  {Goold}}, \bibinfo {author} {\bibfnamefont {M.}~\bibnamefont {Huber}},
  \bibinfo {author} {\bibfnamefont {A.}~\bibnamefont {Riera}}, \bibinfo
  {author} {\bibfnamefont {L.}~\bibnamefont {del Rio}}, \ and\ \bibinfo
  {author} {\bibfnamefont {P.}~\bibnamefont {Skrzypczyk}},\ }\bibfield  {title}
  {\enquote {\bibinfo {title} {The role of quantum information in
  thermodynamics -- a topical review},}\ }\href
  {http://stacks.iop.org/1751-8121/49/i=14/a=143001} {\bibfield  {journal}
  {\bibinfo  {journal} {J. Phys. A: Math. Theor.}\ }\textbf {\bibinfo {volume}
  {49}},\ \bibinfo {pages} {143001} (\bibinfo {year} {2016})}\BibitemShut
  {NoStop}%
\bibitem [{\citenamefont {Vinjanampathy}\ and\ \citenamefont
  {Anders}(2016)}]{vinjanampathy:2016}%
  \BibitemOpen
  \bibfield  {author} {\bibinfo {author} {\bibfnamefont {S.}~\bibnamefont
  {Vinjanampathy}}\ and\ \bibinfo {author} {\bibfnamefont {J.}~\bibnamefont
  {Anders}},\ }\bibfield  {title} {\enquote {\bibinfo {title} {Quantum
  thermodynamics},}\ }\href {\doibase 10.1080/00107514.2016.1201896} {\bibfield
   {journal} {\bibinfo  {journal} {Contemp. Phys.}\ }\textbf {\bibinfo {volume}
  {57}},\ \bibinfo {pages} {545} (\bibinfo {year} {2016})}\BibitemShut
  {NoStop}%
\bibitem [{\citenamefont {Stace}(2010)}]{stace:2010}%
  \BibitemOpen
  \bibfield  {author} {\bibinfo {author} {\bibfnamefont {T.~M.}\ \bibnamefont
  {Stace}},\ }\bibfield  {title} {\enquote {\bibinfo {title} {Quantum limits of
  thermometry},}\ }\href {\doibase 10.1103/PhysRevA.82.011611} {\bibfield
  {journal} {\bibinfo  {journal} {Phys. Rev. A}\ }\textbf {\bibinfo {volume}
  {82}},\ \bibinfo {pages} {011611} (\bibinfo {year} {2010})}\BibitemShut
  {NoStop}%
\bibitem [{\citenamefont {Marzolino}\ and\ \citenamefont
  {Braun}(2013)}]{marzolino:2013}%
  \BibitemOpen
  \bibfield  {author} {\bibinfo {author} {\bibfnamefont {U.}~\bibnamefont
  {Marzolino}}\ and\ \bibinfo {author} {\bibfnamefont {D.}~\bibnamefont
  {Braun}},\ }\bibfield  {title} {\enquote {\bibinfo {title} {Precision
  measurements of temperature and chemical potential of quantum gases},}\
  }\href {\doibase 10.1103/PhysRevA.88.063609} {\bibfield  {journal} {\bibinfo
  {journal} {Phys. Rev. A}\ }\textbf {\bibinfo {volume} {88}},\ \bibinfo
  {pages} {063609} (\bibinfo {year} {2013})}\BibitemShut {NoStop}%
\bibitem [{\citenamefont {Correa}\ \emph {et~al.}(2015)\citenamefont {Correa},
  \citenamefont {Mehboudi}, \citenamefont {Adesso},\ and\ \citenamefont
  {Sanpera}}]{correa:2015}%
  \BibitemOpen
  \bibfield  {author} {\bibinfo {author} {\bibfnamefont {L.~A.}\ \bibnamefont
  {Correa}}, \bibinfo {author} {\bibfnamefont {M.}~\bibnamefont {Mehboudi}},
  \bibinfo {author} {\bibfnamefont {G.}~\bibnamefont {Adesso}}, \ and\ \bibinfo
  {author} {\bibfnamefont {A.}~\bibnamefont {Sanpera}},\ }\bibfield  {title}
  {\enquote {\bibinfo {title} {Individual quantum probes for optimal
  thermometry},}\ }\href {\doibase 10.1103/PhysRevLett.114.220405} {\bibfield
  {journal} {\bibinfo  {journal} {Phys. Rev. Lett.}\ }\textbf {\bibinfo
  {volume} {114}},\ \bibinfo {pages} {220405} (\bibinfo {year}
  {2015})}\BibitemShut {NoStop}%
\bibitem [{\citenamefont {Paris}(2016)}]{paris:2016}%
  \BibitemOpen
  \bibfield  {author} {\bibinfo {author} {\bibfnamefont {M.~G.~A.}\
  \bibnamefont {Paris}},\ }\bibfield  {title} {\enquote {\bibinfo {title}
  {Achieving the {L}andau bound to precision of quantum thermometry in systems
  with vanishing gap},}\ }\href {\doibase 10.1088/1751-8113/49/3/03LT02}
  {\bibfield  {journal} {\bibinfo  {journal} {J. Phys. A: Math. Theor.}\
  }\textbf {\bibinfo {volume} {49}},\ \bibinfo {pages} {03LT02} (\bibinfo
  {year} {2016})}\BibitemShut {NoStop}%
\bibitem [{\citenamefont {De~Pasquale}\ \emph {et~al.}(2016)\citenamefont
  {De~Pasquale}, \citenamefont {Rossini}, \citenamefont {Fazio},\ and\
  \citenamefont {Giovannetti}}]{pasquale:2016}%
  \BibitemOpen
  \bibfield  {author} {\bibinfo {author} {\bibfnamefont {A.}~\bibnamefont
  {De~Pasquale}}, \bibinfo {author} {\bibfnamefont {D.}~\bibnamefont
  {Rossini}}, \bibinfo {author} {\bibfnamefont {R.}~\bibnamefont {Fazio}}, \
  and\ \bibinfo {author} {\bibfnamefont {V.}~\bibnamefont {Giovannetti}},\
  }\bibfield  {title} {\enquote {\bibinfo {title} {Local quantum thermal
  susceptibility},}\ }\href {http://dx.doi.org/10.1038/ncomms12782} {\bibfield
  {journal} {\bibinfo  {journal} {Nat. Commun.}\ }\textbf {\bibinfo {volume}
  {7}},\ \bibinfo {pages} {12782} (\bibinfo {year} {2016})}\BibitemShut
  {NoStop}%
\bibitem [{\citenamefont {Jevtic}\ \emph {et~al.}(2015)\citenamefont {Jevtic},
  \citenamefont {Newman}, \citenamefont {Rudolph},\ and\ \citenamefont
  {Stace}}]{jevtic:2015}%
  \BibitemOpen
  \bibfield  {author} {\bibinfo {author} {\bibfnamefont {S.}~\bibnamefont
  {Jevtic}}, \bibinfo {author} {\bibfnamefont {D.}~\bibnamefont {Newman}},
  \bibinfo {author} {\bibfnamefont {T.}~\bibnamefont {Rudolph}}, \ and\
  \bibinfo {author} {\bibfnamefont {T.~M.}\ \bibnamefont {Stace}},\ }\bibfield
  {title} {\enquote {\bibinfo {title} {Single-qubit thermometry},}\ }\href
  {\doibase 10.1103/PhysRevA.91.012331} {\bibfield  {journal} {\bibinfo
  {journal} {Phys. Rev. A}\ }\textbf {\bibinfo {volume} {91}},\ \bibinfo
  {pages} {012331} (\bibinfo {year} {2015})}\BibitemShut {NoStop}%
\bibitem [{\citenamefont {Johnson}\ \emph {et~al.}(2016)\citenamefont
  {Johnson}, \citenamefont {Cosco}, \citenamefont {Mitchison}, \citenamefont
  {Jaksch},\ and\ \citenamefont {Clark}}]{johnson:2016}%
  \BibitemOpen
  \bibfield  {author} {\bibinfo {author} {\bibfnamefont {T.~H.}\ \bibnamefont
  {Johnson}}, \bibinfo {author} {\bibfnamefont {F.}~\bibnamefont {Cosco}},
  \bibinfo {author} {\bibfnamefont {M.~T.}\ \bibnamefont {Mitchison}}, \bibinfo
  {author} {\bibfnamefont {D.}~\bibnamefont {Jaksch}}, \ and\ \bibinfo {author}
  {\bibfnamefont {S.~R.}\ \bibnamefont {Clark}},\ }\bibfield  {title} {\enquote
  {\bibinfo {title} {Thermometry of ultracold atoms via nonequilibrium work
  distributions},}\ }\href {\doibase 10.1103/PhysRevA.93.053619} {\bibfield
  {journal} {\bibinfo  {journal} {Phys. Rev. A}\ }\textbf {\bibinfo {volume}
  {93}},\ \bibinfo {pages} {053619} (\bibinfo {year} {2016})}\BibitemShut
  {NoStop}%
\bibitem [{\citenamefont {Mart\'in-Mart\'inez}\ \emph
  {et~al.}(2013)\citenamefont {Mart\'in-Mart\'inez}, \citenamefont {Dragan},
  \citenamefont {Mann},\ and\ \citenamefont {Fuentes}}]{martin:2013}%
  \BibitemOpen
  \bibfield  {author} {\bibinfo {author} {\bibfnamefont {E.}~\bibnamefont
  {Mart\'in-Mart\'inez}}, \bibinfo {author} {\bibfnamefont {A.}~\bibnamefont
  {Dragan}}, \bibinfo {author} {\bibfnamefont {R.~B.}\ \bibnamefont {Mann}}, \
  and\ \bibinfo {author} {\bibfnamefont {I.}~\bibnamefont {Fuentes}},\
  }\bibfield  {title} {\enquote {\bibinfo {title} {Berry phase quantum
  thermometer},}\ }\href {http://stacks.iop.org/1367-2630/15/i=5/a=053036}
  {\bibfield  {journal} {\bibinfo  {journal} {New J. Phys.}\ }\textbf {\bibinfo
  {volume} {15}},\ \bibinfo {pages} {053036} (\bibinfo {year}
  {2013})}\BibitemShut {NoStop}%
\bibitem [{\citenamefont {Sab\'in}\ \emph {et~al.}(2014)\citenamefont
  {Sab\'in}, \citenamefont {White}, \citenamefont {Hackermuller},\ and\
  \citenamefont {Fuentes}}]{sabin:2014}%
  \BibitemOpen
  \bibfield  {author} {\bibinfo {author} {\bibfnamefont {C.}~\bibnamefont
  {Sab\'in}}, \bibinfo {author} {\bibfnamefont {A.}~\bibnamefont {White}},
  \bibinfo {author} {\bibfnamefont {L.}~\bibnamefont {Hackermuller}}, \ and\
  \bibinfo {author} {\bibfnamefont {I.}~\bibnamefont {Fuentes}},\ }\bibfield
  {title} {\enquote {\bibinfo {title} {Impurities as a quantum thermometer for
  a {B}ose-{E}instein condensate},}\ }\href
  {http://www.ncbi.nlm.nih.gov/pmc/articles/PMC4170192/} {\bibfield  {journal}
  {\bibinfo  {journal} {Sci. Rep.}\ }\textbf {\bibinfo {volume} {4}},\ \bibinfo
  {pages} {6436} (\bibinfo {year} {2014})}\BibitemShut {NoStop}%
\bibitem [{\citenamefont {Guo}\ \emph {et~al.}(2015)\citenamefont {Guo},
  \citenamefont {Xu}, \citenamefont {Zou},\ and\ \citenamefont
  {Shao}}]{guo:2015}%
  \BibitemOpen
  \bibfield  {author} {\bibinfo {author} {\bibfnamefont {L.-S.}\ \bibnamefont
  {Guo}}, \bibinfo {author} {\bibfnamefont {B.-M.}\ \bibnamefont {Xu}},
  \bibinfo {author} {\bibfnamefont {J.}~\bibnamefont {Zou}}, \ and\ \bibinfo
  {author} {\bibfnamefont {B.}~\bibnamefont {Shao}},\ }\bibfield  {title}
  {\enquote {\bibinfo {title} {Improved thermometry of low-temperature quantum
  systems by a ring-structure probe},}\ }\href {\doibase
  10.1103/PhysRevA.92.052112} {\bibfield  {journal} {\bibinfo  {journal} {Phys.
  Rev. A}\ }\textbf {\bibinfo {volume} {92}},\ \bibinfo {pages} {052112}
  (\bibinfo {year} {2015})}\BibitemShut {NoStop}%
\bibitem [{\citenamefont {Correa}\ \emph {et~al.}(2016)\citenamefont {Correa},
  \citenamefont {Perarnau-Llobet}, \citenamefont {Hovhannisyan}, \citenamefont
  {Hernández-Santana}, \citenamefont {Mehboudi},\ and\ \citenamefont
  {Sanpera}}]{correa:2016}%
  \BibitemOpen
  \bibfield  {author} {\bibinfo {author} {\bibfnamefont {L.~A.}\ \bibnamefont
  {Correa}}, \bibinfo {author} {\bibfnamefont {M.}~\bibnamefont
  {Perarnau-Llobet}}, \bibinfo {author} {\bibfnamefont {K.~V.}\ \bibnamefont
  {Hovhannisyan}}, \bibinfo {author} {\bibfnamefont {S.}~\bibnamefont
  {Hernández-Santana}}, \bibinfo {author} {\bibfnamefont {M.}~\bibnamefont
  {Mehboudi}}, \ and\ \bibinfo {author} {\bibfnamefont {A.}~\bibnamefont
  {Sanpera}},\ }\href@noop {} {\enquote {\bibinfo {title} {Low-temperature
  thermometry enhanced by strong coupling},}\ } (\bibinfo {year} {2016}),\
  \Eprint {http://arxiv.org/abs/1611.10123} {arXiv:1611.10123 [quant-ph]}
  \BibitemShut {NoStop}%
\bibitem [{\citenamefont {Giazotto}\ \emph {et~al.}(2006)\citenamefont
  {Giazotto}, \citenamefont {Heikkil\"a}, \citenamefont {Luukanen},
  \citenamefont {Savin},\ and\ \citenamefont {Pekola}}]{giazotto:2006}%
  \BibitemOpen
  \bibfield  {author} {\bibinfo {author} {\bibfnamefont {F.}~\bibnamefont
  {Giazotto}}, \bibinfo {author} {\bibfnamefont {T.~T.}\ \bibnamefont
  {Heikkil\"a}}, \bibinfo {author} {\bibfnamefont {A.}~\bibnamefont
  {Luukanen}}, \bibinfo {author} {\bibfnamefont {A.~M.}\ \bibnamefont {Savin}},
  \ and\ \bibinfo {author} {\bibfnamefont {J.~P.}\ \bibnamefont {Pekola}},\
  }\bibfield  {title} {\enquote {\bibinfo {title} {Opportunities for
  mesoscopics in thermometry and refrigeration: Physics and applications},}\
  }\href {\doibase 10.1103/RevModPhys.78.217} {\bibfield  {journal} {\bibinfo
  {journal} {Rev. Mod. Phys.}\ }\textbf {\bibinfo {volume} {78}},\ \bibinfo
  {pages} {217} (\bibinfo {year} {2006})}\BibitemShut {NoStop}%
\bibitem [{\citenamefont {Iftikhar}\ \emph {et~al.}(2016)\citenamefont
  {Iftikhar}, \citenamefont {Anthore}, \citenamefont {Jezouin}, \citenamefont
  {Parmentier}, \citenamefont {Jin}, \citenamefont {Cavanna}, \citenamefont
  {Ouerghi}, \citenamefont {Gennser},\ and\ \citenamefont
  {Pierre}}]{iftikhar:2016}%
  \BibitemOpen
  \bibfield  {author} {\bibinfo {author} {\bibfnamefont {Z.}~\bibnamefont
  {Iftikhar}}, \bibinfo {author} {\bibfnamefont {A.}~\bibnamefont {Anthore}},
  \bibinfo {author} {\bibfnamefont {S.}~\bibnamefont {Jezouin}}, \bibinfo
  {author} {\bibfnamefont {F.~D.}\ \bibnamefont {Parmentier}}, \bibinfo
  {author} {\bibfnamefont {Y.}~\bibnamefont {Jin}}, \bibinfo {author}
  {\bibfnamefont {A.}~\bibnamefont {Cavanna}}, \bibinfo {author} {\bibfnamefont
  {A.}~\bibnamefont {Ouerghi}}, \bibinfo {author} {\bibfnamefont
  {U.}~\bibnamefont {Gennser}}, \ and\ \bibinfo {author} {\bibfnamefont
  {F.}~\bibnamefont {Pierre}},\ }\bibfield  {title} {\enquote {\bibinfo {title}
  {Primary thermometry triad at 6 m{K} in mesoscopic circuits},}\ }\href
  {http://dx.doi.org/10.1038/ncomms12908} {\bibfield  {journal} {\bibinfo
  {journal} {Nat. Commun.}\ }\textbf {\bibinfo {volume} {7}},\ \bibinfo {pages}
  {12908} (\bibinfo {year} {2016})}\BibitemShut {NoStop}%
\bibitem [{\citenamefont {Zgirski}\ \emph {et~al.}(2017)\citenamefont
  {Zgirski}, \citenamefont {Foltyn},\ and\ \citenamefont
  {Pekola}}]{zgirski:2017}%
  \BibitemOpen
  \bibfield  {author} {\bibinfo {author} {\bibfnamefont {M.}~\bibnamefont
  {Zgirski}}, \bibinfo {author} {\bibfnamefont {M.}~\bibnamefont {Foltyn}}, \
  and\ \bibinfo {author} {\bibfnamefont {A.~Savinnd M. Meschkeand~J.}\
  \bibnamefont {Pekola}},\ }\href@noop {} {\enquote {\bibinfo {title}
  {Nanosecond thermometry with josephson junction},}\ } (\bibinfo {year}
  {2017}),\ \Eprint {http://arxiv.org/abs/1704.04762} {arXiv:1704.04762
  [cond-mat]} \BibitemShut {NoStop}%
\bibitem [{\citenamefont {Benenti}\ \emph {et~al.}(2017)\citenamefont
  {Benenti}, \citenamefont {Casati}, \citenamefont {Saito},\ and\ \citenamefont
  {Whitney}}]{benenti:2017}%
  \BibitemOpen
  \bibfield  {author} {\bibinfo {author} {\bibfnamefont {G.}~\bibnamefont
  {Benenti}}, \bibinfo {author} {\bibfnamefont {G.}~\bibnamefont {Casati}},
  \bibinfo {author} {\bibfnamefont {K.}~\bibnamefont {Saito}}, \ and\ \bibinfo
  {author} {\bibfnamefont {R. S.}\ \bibnamefont {Whitney}},\ }\bibfield
  {title} {\enquote {\bibinfo {title} {Fundamental aspects of steady-state
  conversion of heat to work at the nanoscale},}\ }\href {\doibase
  10.1016/j.physrep.2017.05.008} {\bibfield  {journal} {\bibinfo  {journal}
  {Phys. Rep.}\ ,\ \bibinfo {pages} {--}} (\bibinfo {year} {2017})}\BibitemShut
  {NoStop}%
\bibitem [{\citenamefont {Kosloff}\ and\ \citenamefont
  {Levy}(2014)}]{kosloff:2014}%
  \BibitemOpen
  \bibfield  {author} {\bibinfo {author} {\bibfnamefont {R.}~\bibnamefont
  {Kosloff}}\ and\ \bibinfo {author} {\bibfnamefont {A.}~\bibnamefont {Levy}},\
  }\bibfield  {title} {\enquote {\bibinfo {title} {Quantum heat engines and
  refrigerators: Continuous devices},}\ }\href {\doibase
  10.1146/annurev-physchem-040513-103724} {\bibfield  {journal} {\bibinfo
  {journal} {Annu. Rev. Phys. Chem.}\ }\textbf {\bibinfo {volume} {65}},\
  \bibinfo {pages} {365} (\bibinfo {year} {2014})}\BibitemShut {NoStop}%
\bibitem [{\citenamefont {Quan}\ \emph {et~al.}(2007)\citenamefont {Quan},
  \citenamefont {Liu}, \citenamefont {Sun},\ and\ \citenamefont
  {Nori}}]{quan:2007}%
  \BibitemOpen
  \bibfield  {author} {\bibinfo {author} {\bibfnamefont {H.~T.}\ \bibnamefont
  {Quan}}, \bibinfo {author} {\bibfnamefont {Yu-xi}\ \bibnamefont {Liu}},
  \bibinfo {author} {\bibfnamefont {C.~P.}\ \bibnamefont {Sun}}, \ and\
  \bibinfo {author} {\bibfnamefont {Franco}\ \bibnamefont {Nori}},\ }\bibfield
  {title} {\enquote {\bibinfo {title} {Quantum thermodynamic cycles and quantum
  heat engines},}\ }\href {\doibase 10.1103/PhysRevE.76.031105} {\bibfield
  {journal} {\bibinfo  {journal} {Phys. Rev. E}\ }\textbf {\bibinfo {volume}
  {76}},\ \bibinfo {pages} {031105} (\bibinfo {year} {2007})}\BibitemShut
  {NoStop}%
\bibitem [{\citenamefont {Brunner}\ \emph {et~al.}(2012)\citenamefont
  {Brunner}, \citenamefont {Linden}, \citenamefont {Popescu},\ and\
  \citenamefont {Skrzypczyk}}]{brunner:2012}%
  \BibitemOpen
  \bibfield  {author} {\bibinfo {author} {\bibfnamefont {N.}~\bibnamefont
  {Brunner}}, \bibinfo {author} {\bibfnamefont {N.}~\bibnamefont {Linden}},
  \bibinfo {author} {\bibfnamefont {S.}~\bibnamefont {Popescu}}, \ and\
  \bibinfo {author} {\bibfnamefont {P.}~\bibnamefont {Skrzypczyk}},\ }\bibfield
   {title} {\enquote {\bibinfo {title} {Virtual qubits, virtual temperatures,
  and the foundations of thermodynamics},}\ }\href {\doibase
  10.1103/PhysRevE.85.051117} {\bibfield  {journal} {\bibinfo  {journal} {Phys.
  Rev. E}\ }\textbf {\bibinfo {volume} {85}},\ \bibinfo {pages} {051117}
  (\bibinfo {year} {2012})}\BibitemShut {NoStop}%
\bibitem [{\citenamefont {Thomson}(1848)}]{kelvin:1848}%
  \BibitemOpen
  \bibfield  {author} {\bibinfo {author} {\bibfnamefont {W.}~\bibnamefont
  {Thomson}},\ }\bibfield  {title} {\enquote {\bibinfo {title} {On an absolute
  thermometric scale founded on {C}arnot's theory of the motive power of heat,
  and calculated from {R}egnault's observations},}\ }\href@noop {} {\bibfield
  {journal} {\bibinfo  {journal} {Proc. Camb. Philos. Soc.}\ }\textbf {\bibinfo
  {volume} {1}},\ \bibinfo {pages} {66} (\bibinfo {year} {1848})}\BibitemShut
  {NoStop}%
\bibitem [{\citenamefont {Geusic}\ \emph {et~al.}(1967)\citenamefont {Geusic},
  \citenamefont {Schulz-DuBois},\ and\ \citenamefont {Scovil}}]{geusic:1967}%
  \BibitemOpen
  \bibfield  {author} {\bibinfo {author} {\bibfnamefont {J.~E.}\ \bibnamefont
  {Geusic}}, \bibinfo {author} {\bibfnamefont {E.~O.}\ \bibnamefont
  {Schulz-DuBois}}, \ and\ \bibinfo {author} {\bibfnamefont {H.~E.~D.}\
  \bibnamefont {Scovil}},\ }\bibfield  {title} {\enquote {\bibinfo {title}
  {Quantum equivalent of the {C}arnot cycle},}\ }\href {\doibase
  10.1103/PhysRev.156.343} {\bibfield  {journal} {\bibinfo  {journal} {Phys.
  Rev.}\ }\textbf {\bibinfo {volume} {156}},\ \bibinfo {pages} {343} (\bibinfo
  {year} {1967})}\BibitemShut {NoStop}%
\bibitem [{\citenamefont {Linden}\ \emph {et~al.}(2010)\citenamefont {Linden},
  \citenamefont {Popescu},\ and\ \citenamefont {Skrzypczyk}}]{linden:2010prl}%
  \BibitemOpen
  \bibfield  {author} {\bibinfo {author} {\bibfnamefont {N.}~\bibnamefont
  {Linden}}, \bibinfo {author} {\bibfnamefont {S.}~\bibnamefont {Popescu}}, \
  and\ \bibinfo {author} {\bibfnamefont {P.}~\bibnamefont {Skrzypczyk}},\
  }\bibfield  {title} {\enquote {\bibinfo {title} {How small can thermal
  machines be? the smallest possible refrigerator},}\ }\href {\doibase
  10.1103/PhysRevLett.105.130401} {\bibfield  {journal} {\bibinfo  {journal}
  {Phys. Rev. Lett.}\ }\textbf {\bibinfo {volume} {105}},\ \bibinfo {pages}
  {130401} (\bibinfo {year} {2010})}\BibitemShut {NoStop}%
\bibitem [{\citenamefont {Skrzypczyk}\ \emph {et~al.}(2011)\citenamefont
  {Skrzypczyk}, \citenamefont {Brunner}, \citenamefont {Linden},\ and\
  \citenamefont {Popescu}}]{skrzypczyk:2011}%
  \BibitemOpen
  \bibfield  {author} {\bibinfo {author} {\bibfnamefont {P.}~\bibnamefont
  {Skrzypczyk}}, \bibinfo {author} {\bibfnamefont {N.}~\bibnamefont {Brunner}},
  \bibinfo {author} {\bibfnamefont {N.}~\bibnamefont {Linden}}, \ and\ \bibinfo
  {author} {\bibfnamefont {S.}~\bibnamefont {Popescu}},\ }\bibfield  {title}
  {\enquote {\bibinfo {title} {The smallest refrigerators can reach maximal
  efficiency},}\ }\href {http://stacks.iop.org/1751-8121/44/i=49/a=492002}
  {\bibfield  {journal} {\bibinfo  {journal} {J. Phys. A}\ }\textbf {\bibinfo
  {volume} {44}},\ \bibinfo {pages} {492002} (\bibinfo {year}
  {2011})}\BibitemShut {NoStop}%
\bibitem [{\citenamefont {Hofer}\ \emph
  {et~al.}(2016{\natexlab{a}})\citenamefont {Hofer}, \citenamefont
  {Perarnau-Llobet}, \citenamefont {Brask}, \citenamefont {Silva},
  \citenamefont {Huber},\ and\ \citenamefont {Brunner}}]{hofer:2016}%
  \BibitemOpen
  \bibfield  {author} {\bibinfo {author} {\bibfnamefont {P.~P.}\ \bibnamefont
  {Hofer}}, \bibinfo {author} {\bibfnamefont {M.}~\bibnamefont
  {Perarnau-Llobet}}, \bibinfo {author} {\bibfnamefont {J.~B.}\ \bibnamefont
  {Brask}}, \bibinfo {author} {\bibfnamefont {R.}~\bibnamefont {Silva}},
  \bibinfo {author} {\bibfnamefont {M.}~\bibnamefont {Huber}}, \ and\ \bibinfo
  {author} {\bibfnamefont {N.}~\bibnamefont {Brunner}},\ }\bibfield  {title}
  {\enquote {\bibinfo {title} {Autonomous quantum refrigerator in a circuit
  {QED} architecture based on a {J}osephson junction},}\ }\href {\doibase
  10.1103/PhysRevB.94.235420} {\bibfield  {journal} {\bibinfo  {journal} {Phys.
  Rev. B}\ }\textbf {\bibinfo {volume} {94}},\ \bibinfo {pages} {235420}
  (\bibinfo {year} {2016}{\natexlab{a}})}\BibitemShut {NoStop}%
\bibitem [{\citenamefont {Mitchison}\ \emph {et~al.}(2016)\citenamefont
  {Mitchison}, \citenamefont {Huber}, \citenamefont {Prior}, \citenamefont
  {Woods},\ and\ \citenamefont {Plenio}}]{mitchison:2016}%
  \BibitemOpen
  \bibfield  {author} {\bibinfo {author} {\bibfnamefont {M.~T.}\ \bibnamefont
  {Mitchison}}, \bibinfo {author} {\bibfnamefont {M.}~\bibnamefont {Huber}},
  \bibinfo {author} {\bibfnamefont {J.}~\bibnamefont {Prior}}, \bibinfo
  {author} {\bibfnamefont {M.~P.}\ \bibnamefont {Woods}}, \ and\ \bibinfo
  {author} {\bibfnamefont {M.~B.}\ \bibnamefont {Plenio}},\ }\bibfield  {title}
  {\enquote {\bibinfo {title} {Realising a quantum absorption refrigerator with
  an atom-cavity system},}\ }\href {\doibase 10.1088/2058-9565/1/1/015001}
  {\bibfield  {journal} {\bibinfo  {journal} {Quantum Sci. Technol.}\ }\textbf
  {\bibinfo {volume} {1}},\ \bibinfo {pages} {015001} (\bibinfo {year}
  {2016})}\BibitemShut {NoStop}%
\bibitem [{\citenamefont {Levy}\ and\ \citenamefont
  {Kosloff}(2012)}]{levy:2012}%
  \BibitemOpen
  \bibfield  {author} {\bibinfo {author} {\bibfnamefont {A.}~\bibnamefont
  {Levy}}\ and\ \bibinfo {author} {\bibfnamefont {R.}~\bibnamefont {Kosloff}},\
  }\bibfield  {title} {\enquote {\bibinfo {title} {Quantum absorption
  refrigerator},}\ }\href {\doibase 10.1103/PhysRevLett.108.070604} {\bibfield
  {journal} {\bibinfo  {journal} {Phys. Rev. Lett.}\ }\textbf {\bibinfo
  {volume} {108}},\ \bibinfo {pages} {070604} (\bibinfo {year}
  {2012})}\BibitemShut {NoStop}%
\bibitem [{\citenamefont {Mitchison}\ \emph {et~al.}(2015)\citenamefont
  {Mitchison}, \citenamefont {Woods}, \citenamefont {P.},\ and\ \citenamefont
  {Huber}}]{mitchison:2015}%
  \BibitemOpen
  \bibfield  {author} {\bibinfo {author} {\bibfnamefont {M.~T.}\ \bibnamefont
  {Mitchison}}, \bibinfo {author} {\bibfnamefont {M.~P.}\ \bibnamefont
  {Woods}}, \bibinfo {author} {\bibfnamefont {J.}~\bibnamefont {P.}}, \ and\
  \bibinfo {author} {\bibfnamefont {M.}~\bibnamefont {Huber}},\ }\bibfield
  {title} {\enquote {\bibinfo {title} {Coherence-assisted single-shot cooling
  by quantum absorption refrigerators},}\ }\href
  {http://stacks.iop.org/1367-2630/17/i=11/a=115013} {\bibfield  {journal}
  {\bibinfo  {journal} {New J. Phys.}\ }\textbf {\bibinfo {volume} {17}},\
  \bibinfo {pages} {115013} (\bibinfo {year} {2015})}\BibitemShut {NoStop}%
\bibitem [{\citenamefont {Silva}\ \emph {et~al.}(2015)\citenamefont {Silva},
  \citenamefont {Skrzypczyk},\ and\ \citenamefont {Brunner}}]{silva:2015}%
  \BibitemOpen
  \bibfield  {author} {\bibinfo {author} {\bibfnamefont {R.}~\bibnamefont
  {Silva}}, \bibinfo {author} {\bibfnamefont {P.}~\bibnamefont {Skrzypczyk}}, \
  and\ \bibinfo {author} {\bibfnamefont {N.}~\bibnamefont {Brunner}},\
  }\bibfield  {title} {\enquote {\bibinfo {title} {Small quantum absorption
  refrigerator with reversed couplings},}\ }\href {\doibase
  10.1103/PhysRevE.92.012136} {\bibfield  {journal} {\bibinfo  {journal} {Phys.
  Rev. E}\ }\textbf {\bibinfo {volume} {92}},\ \bibinfo {pages} {012136}
  (\bibinfo {year} {2015})}\BibitemShut {NoStop}%
\bibitem [{\citenamefont {Brunner}\ \emph {et~al.}(2014)\citenamefont
  {Brunner}, \citenamefont {Huber}, \citenamefont {Linden}, \citenamefont
  {Popescu}, \citenamefont {Silva},\ and\ \citenamefont
  {Skrzypczyk}}]{brunner:2014}%
  \BibitemOpen
  \bibfield  {author} {\bibinfo {author} {\bibfnamefont {N.}~\bibnamefont
  {Brunner}}, \bibinfo {author} {\bibfnamefont {M.}~\bibnamefont {Huber}},
  \bibinfo {author} {\bibfnamefont {N.}~\bibnamefont {Linden}}, \bibinfo
  {author} {\bibfnamefont {S.}~\bibnamefont {Popescu}}, \bibinfo {author}
  {\bibfnamefont {R.}~\bibnamefont {Silva}}, \ and\ \bibinfo {author}
  {\bibfnamefont {P.}~\bibnamefont {Skrzypczyk}},\ }\bibfield  {title}
  {\enquote {\bibinfo {title} {Entanglement enhances cooling in microscopic
  quantum refrigerators},}\ }\href {\doibase 10.1103/PhysRevE.89.032115}
  {\bibfield  {journal} {\bibinfo  {journal} {Phys. Rev. E}\ }\textbf {\bibinfo
  {volume} {89}},\ \bibinfo {pages} {032115} (\bibinfo {year}
  {2014})}\BibitemShut {NoStop}%
\bibitem [{\citenamefont {Brask}\ and\ \citenamefont
  {Brunner}(2015)}]{brask:2015}%
  \BibitemOpen
  \bibfield  {author} {\bibinfo {author} {\bibfnamefont {J.~B.}\ \bibnamefont
  {Brask}}\ and\ \bibinfo {author} {\bibfnamefont {N.}~\bibnamefont
  {Brunner}},\ }\bibfield  {title} {\enquote {\bibinfo {title} {Small quantum
  absorption refrigerator in the transient regime: Time scales, enhanced
  cooling, and entanglement},}\ }\href {\doibase 10.1103/PhysRevE.92.062101}
  {\bibfield  {journal} {\bibinfo  {journal} {Phys. Rev. E}\ }\textbf {\bibinfo
  {volume} {92}},\ \bibinfo {pages} {062101} (\bibinfo {year}
  {2015})}\BibitemShut {NoStop}%
\bibitem [{\citenamefont {Roulet}\ \emph {et~al.}(2017)\citenamefont {Roulet},
  \citenamefont {Nimmrichter}, \citenamefont {Arrazola}, \citenamefont {Seah},\
  and\ \citenamefont {Scarani}}]{roulet:2017}%
  \BibitemOpen
  \bibfield  {author} {\bibinfo {author} {\bibfnamefont {A.}~\bibnamefont
  {Roulet}}, \bibinfo {author} {\bibfnamefont {S.}~\bibnamefont {Nimmrichter}},
  \bibinfo {author} {\bibfnamefont {J.~M.}\ \bibnamefont {Arrazola}}, \bibinfo
  {author} {\bibfnamefont {S.}~\bibnamefont {Seah}}, \ and\ \bibinfo {author}
  {\bibfnamefont {V.}~\bibnamefont {Scarani}},\ }\bibfield  {title} {\enquote
  {\bibinfo {title} {Autonomous rotor heat engine},}\ }\href {\doibase
  10.1103/PhysRevE.95.062131} {\bibfield  {journal} {\bibinfo  {journal} {Phys.
  Rev. E}\ }\textbf {\bibinfo {volume} {95}},\ \bibinfo {pages} {062131}
  (\bibinfo {year} {2017})}\BibitemShut {NoStop}%
\bibitem [{sup()}]{supplementarx}%
  \BibitemOpen
  \href@noop {} {}\bibinfo {note} {See supplemental Material below for
  analytical results using a simplified model and a discussion on the influence
  of heat leaks and spurious modes.}\BibitemShut {Stop}%
\bibitem [{mah(2009)}]{mahler:book}%
  \BibitemOpen
  \href {\doibase 10.1007/978-3-540-70510-9} {\emph {\bibinfo {title} {Quantum
  Thermodynamics: Emergence of Thermodynamic Behavior Within Composite Quantum
  Systems}}},\ \bibinfo {series} {Lecture Notes in Physics}, Vol.\ \bibinfo
  {volume} {784}\ (\bibinfo  {publisher} {Springer (Berlin)},\ \bibinfo {year}
  {2009})\BibitemShut {NoStop}%
\bibitem [{\citenamefont {Kosloff}(1984)}]{kosloff:1984}%
  \BibitemOpen
  \bibfield  {author} {\bibinfo {author} {\bibfnamefont {R.}~\bibnamefont
  {Kosloff}},\ }\bibfield  {title} {\enquote {\bibinfo {title} {A quantum
  mechanical open system as a model of a heat engine},}\ }\href {\doibase
  10.1063/1.446862} {\bibfield  {journal} {\bibinfo  {journal} {J. Chem.
  Phys.}\ }\textbf {\bibinfo {volume} {80}},\ \bibinfo {pages} {1625} (\bibinfo
  {year} {1984})}\BibitemShut {NoStop}%
\bibitem [{\citenamefont {Feldmann}\ and\ \citenamefont
  {Kosloff}(2003)}]{feldmann:2003}%
  \BibitemOpen
  \bibfield  {author} {\bibinfo {author} {\bibfnamefont {T.}~\bibnamefont
  {Feldmann}}\ and\ \bibinfo {author} {\bibfnamefont {R.}~\bibnamefont
  {Kosloff}},\ }\bibfield  {title} {\enquote {\bibinfo {title} {Quantum
  four-stroke heat engine: {T}hermodynamic observables in a model with
  intrinsic friction},}\ }\href {\doibase 10.1103/PhysRevE.68.016101}
  {\bibfield  {journal} {\bibinfo  {journal} {Phys. Rev. E}\ }\textbf {\bibinfo
  {volume} {68}},\ \bibinfo {pages} {016101} (\bibinfo {year}
  {2003})}\BibitemShut {NoStop}%
\bibitem [{\citenamefont {Niskanen}\ \emph {et~al.}(2007)\citenamefont
  {Niskanen}, \citenamefont {Nakamura},\ and\ \citenamefont
  {Pekola}}]{niskanen:2007}%
  \BibitemOpen
  \bibfield  {author} {\bibinfo {author} {\bibfnamefont {A.~O.}\ \bibnamefont
  {Niskanen}}, \bibinfo {author} {\bibfnamefont {Y.}~\bibnamefont {Nakamura}},
  \ and\ \bibinfo {author} {\bibfnamefont {J.~P.}\ \bibnamefont {Pekola}},\
  }\bibfield  {title} {\enquote {\bibinfo {title} {Information entropic
  superconducting microcooler},}\ }\href {\doibase 10.1103/PhysRevB.76.174523}
  {\bibfield  {journal} {\bibinfo  {journal} {Phys. Rev. B}\ }\textbf {\bibinfo
  {volume} {76}},\ \bibinfo {pages} {174523} (\bibinfo {year}
  {2007})}\BibitemShut {NoStop}%
\bibitem [{\citenamefont {Henrich}\ \emph {et~al.}(2007)\citenamefont
  {Henrich}, \citenamefont {Rempp},\ and\ \citenamefont
  {Mahler}}]{henrich:2007}%
  \BibitemOpen
  \bibfield  {author} {\bibinfo {author} {\bibfnamefont {M.~J.}\ \bibnamefont
  {Henrich}}, \bibinfo {author} {\bibfnamefont {F.}~\bibnamefont {Rempp}}, \
  and\ \bibinfo {author} {\bibfnamefont {G.}~\bibnamefont {Mahler}},\
  }\bibfield  {title} {\enquote {\bibinfo {title} {Quantum thermodynamic {O}tto
  machines: A spin-system approach},}\ }\href {\doibase
  10.1140/epjst/e2007-00371-8} {\bibfield  {journal} {\bibinfo  {journal} {Eur.
  Phys. J. ST}\ }\textbf {\bibinfo {volume} {151}},\ \bibinfo {pages} {157}
  (\bibinfo {year} {2007})}\BibitemShut {NoStop}%
\bibitem [{\citenamefont {Hofer}\ \emph
  {et~al.}(2016{\natexlab{b}})\citenamefont {Hofer}, \citenamefont {Souquet},\
  and\ \citenamefont {Clerk}}]{hofer:2016prb}%
  \BibitemOpen
  \bibfield  {author} {\bibinfo {author} {\bibfnamefont {P.~P.}\ \bibnamefont
  {Hofer}}, \bibinfo {author} {\bibfnamefont {J.-R.}\ \bibnamefont {Souquet}},
  \ and\ \bibinfo {author} {\bibfnamefont {A.~A.}\ \bibnamefont {Clerk}},\
  }\bibfield  {title} {\enquote {\bibinfo {title} {Quantum heat engine based on
  photon-assisted {C}ooper pair tunneling},}\ }\href {\doibase
  10.1103/PhysRevB.93.041418} {\bibfield  {journal} {\bibinfo  {journal} {Phys.
  Rev. B}\ }\textbf {\bibinfo {volume} {93}},\ \bibinfo {pages} {041418(R)}
  (\bibinfo {year} {2016}{\natexlab{b}})}\BibitemShut {NoStop}%
\bibitem [{\citenamefont {Campisi}\ and\ \citenamefont
  {Fazio}(2016)}]{campisi:2016}%
  \BibitemOpen
  \bibfield  {author} {\bibinfo {author} {\bibfnamefont {M.}~\bibnamefont
  {Campisi}}\ and\ \bibinfo {author} {\bibfnamefont {R.}~\bibnamefont
  {Fazio}},\ }\bibfield  {title} {\enquote {\bibinfo {title} {The power of a
  critical heat engine},}\ }\href {http://dx.doi.org/10.1038/ncomms11895}
  {\bibfield  {journal} {\bibinfo  {journal} {Nat. Commun.}\ }\textbf {\bibinfo
  {volume} {7}},\ \bibinfo {pages} {11895} (\bibinfo {year}
  {2016})}\BibitemShut {NoStop}%
\bibitem [{\citenamefont {Westig}\ \emph {et~al.}(2017)\citenamefont {Westig},
  \citenamefont {Kubala}, \citenamefont {Parlavecchio}, \citenamefont
  {Mukharsky}, \citenamefont {Altimiras}, \citenamefont {Joyez}, \citenamefont
  {Vion}, \citenamefont {Roche}, \citenamefont {Hofheinz}, \citenamefont
  {Esteve}, \citenamefont {Trif}, \citenamefont {Simon}, \citenamefont
  {Ankerhold},\ and\ \citenamefont {Portier}}]{westig:2017}%
  \BibitemOpen
  \bibfield  {author} {\bibinfo {author} {\bibfnamefont {M.}~\bibnamefont
  {Westig}}, \bibinfo {author} {\bibfnamefont {B.}~\bibnamefont {Kubala}},
  \bibinfo {author} {\bibfnamefont {O.}~\bibnamefont {Parlavecchio}}, \bibinfo
  {author} {\bibfnamefont {Y.}~\bibnamefont {Mukharsky}}, \bibinfo {author}
  {\bibfnamefont {C.}~\bibnamefont {Altimiras}}, \bibinfo {author}
  {\bibfnamefont {P}~\bibnamefont {Joyez}}, \bibinfo {author} {\bibfnamefont
  {D.}~\bibnamefont {Vion}}, \bibinfo {author} {\bibfnamefont {P.}~\bibnamefont
  {Roche}}, \bibinfo {author} {\bibfnamefont {M.}~\bibnamefont {Hofheinz}},
  \bibinfo {author} {\bibfnamefont {D}~\bibnamefont {Esteve}}, \bibinfo
  {author} {\bibfnamefont {M}~\bibnamefont {Trif}}, \bibinfo {author}
  {\bibfnamefont {P.}~\bibnamefont {Simon}}, \bibinfo {author} {\bibfnamefont
  {J.}~\bibnamefont {Ankerhold}}, \ and\ \bibinfo {author} {\bibfnamefont
  {F.}~\bibnamefont {Portier}},\ }\href@noop {} {\enquote {\bibinfo {title}
  {Emission of non-classical radiation by inelastic cooper pair tunneling},}\ }
  (\bibinfo {year} {2017}),\ \Eprint {http://arxiv.org/abs/1703.05009}
  {arXiv:1703.05009 [cond-mat]} \BibitemShut {NoStop}%
\bibitem [{\citenamefont {Holst}\ \emph {et~al.}(1994)\citenamefont {Holst},
  \citenamefont {Esteve}, \citenamefont {Urbina},\ and\ \citenamefont
  {Devoret}}]{holst:1994}%
  \BibitemOpen
  \bibfield  {author} {\bibinfo {author} {\bibfnamefont {T.}~\bibnamefont
  {Holst}}, \bibinfo {author} {\bibfnamefont {D.}~\bibnamefont {Esteve}},
  \bibinfo {author} {\bibfnamefont {C.}~\bibnamefont {Urbina}}, \ and\ \bibinfo
  {author} {\bibfnamefont {M.~H.}\ \bibnamefont {Devoret}},\ }\bibfield
  {title} {\enquote {\bibinfo {title} {Effect of a transmission line resonator
  on a small capacitance tunnel junction},}\ }\href {\doibase
  10.1103/PhysRevLett.73.3455} {\bibfield  {journal} {\bibinfo  {journal}
  {Phys. Rev. Lett.}\ }\textbf {\bibinfo {volume} {73}},\ \bibinfo {pages}
  {3455} (\bibinfo {year} {1994})}\BibitemShut {NoStop}%
\bibitem [{\citenamefont {Basset}\ \emph {et~al.}(2010)\citenamefont {Basset},
  \citenamefont {Bouchiat},\ and\ \citenamefont {Deblock}}]{basset:2010}%
  \BibitemOpen
  \bibfield  {author} {\bibinfo {author} {\bibfnamefont {J.}~\bibnamefont
  {Basset}}, \bibinfo {author} {\bibfnamefont {H.}~\bibnamefont {Bouchiat}}, \
  and\ \bibinfo {author} {\bibfnamefont {R.}~\bibnamefont {Deblock}},\
  }\bibfield  {title} {\enquote {\bibinfo {title} {Emission and absorption
  quantum noise measurement with an on-chip resonant circuit},}\ }\href
  {\doibase 10.1103/PhysRevLett.105.166801} {\bibfield  {journal} {\bibinfo
  {journal} {Phys. Rev. Lett.}\ }\textbf {\bibinfo {volume} {105}},\ \bibinfo
  {pages} {166801} (\bibinfo {year} {2010})}\BibitemShut {NoStop}%
\bibitem [{\citenamefont {Hofheinz}\ \emph {et~al.}(2011)\citenamefont
  {Hofheinz}, \citenamefont {Portier}, \citenamefont {Baudouin}, \citenamefont
  {Joyez}, \citenamefont {Vion}, \citenamefont {Bertet}, \citenamefont
  {Roche},\ and\ \citenamefont {Esteve}}]{hofheinz:2011}%
  \BibitemOpen
  \bibfield  {author} {\bibinfo {author} {\bibfnamefont {M.}~\bibnamefont
  {Hofheinz}}, \bibinfo {author} {\bibfnamefont {F.}~\bibnamefont {Portier}},
  \bibinfo {author} {\bibfnamefont {Q.}~\bibnamefont {Baudouin}}, \bibinfo
  {author} {\bibfnamefont {P.}~\bibnamefont {Joyez}}, \bibinfo {author}
  {\bibfnamefont {D.}~\bibnamefont {Vion}}, \bibinfo {author} {\bibfnamefont
  {P.}~\bibnamefont {Bertet}}, \bibinfo {author} {\bibfnamefont
  {P.}~\bibnamefont {Roche}}, \ and\ \bibinfo {author} {\bibfnamefont
  {D.}~\bibnamefont {Esteve}},\ }\bibfield  {title} {\enquote {\bibinfo {title}
  {Bright side of the {C}oulomb blockade},}\ }\href {\doibase
  10.1103/PhysRevLett.106.217005} {\bibfield  {journal} {\bibinfo  {journal}
  {Phys. Rev. Lett.}\ }\textbf {\bibinfo {volume} {106}},\ \bibinfo {pages}
  {217005} (\bibinfo {year} {2011})}\BibitemShut {NoStop}%
\bibitem [{\citenamefont {Saira}\ \emph {et~al.}(2016)\citenamefont {Saira},
  \citenamefont {Zgirski}, \citenamefont {Viisanen}, \citenamefont {Golubev},\
  and\ \citenamefont {Pekola}}]{saira:2016}%
  \BibitemOpen
  \bibfield  {author} {\bibinfo {author} {\bibfnamefont {O.-P.}\ \bibnamefont
  {Saira}}, \bibinfo {author} {\bibfnamefont {M.}~\bibnamefont {Zgirski}},
  \bibinfo {author} {\bibfnamefont {K.~L.}\ \bibnamefont {Viisanen}}, \bibinfo
  {author} {\bibfnamefont {D.~S.}\ \bibnamefont {Golubev}}, \ and\ \bibinfo
  {author} {\bibfnamefont {J.~P.}\ \bibnamefont {Pekola}},\ }\bibfield  {title}
  {\enquote {\bibinfo {title} {Dispersive thermometry with a {J}osephson
  junction coupled to a resonator},}\ }\href {\doibase
  10.1103/PhysRevApplied.6.024005} {\bibfield  {journal} {\bibinfo  {journal}
  {Phys. Rev. Applied}\ }\textbf {\bibinfo {volume} {6}},\ \bibinfo {pages}
  {024005} (\bibinfo {year} {2016})}\BibitemShut {NoStop}%
\bibitem [{\citenamefont {Gramich}\ \emph {et~al.}(2013)\citenamefont
  {Gramich}, \citenamefont {Kubala}, \citenamefont {Rohrer},\ and\
  \citenamefont {Ankerhold}}]{gramich:2013}%
  \BibitemOpen
  \bibfield  {author} {\bibinfo {author} {\bibfnamefont {V.}~\bibnamefont
  {Gramich}}, \bibinfo {author} {\bibfnamefont {B.}~\bibnamefont {Kubala}},
  \bibinfo {author} {\bibfnamefont {S.}~\bibnamefont {Rohrer}}, \ and\ \bibinfo
  {author} {\bibfnamefont {J.}~\bibnamefont {Ankerhold}},\ }\bibfield  {title}
  {\enquote {\bibinfo {title} {From {C}oulomb-blockade to nonlinear quantum
  dynamics in a superconducting circuit with a resonator},}\ }\href {\doibase
  10.1103/PhysRevLett.111.247002} {\bibfield  {journal} {\bibinfo  {journal}
  {Phys. Rev. Lett.}\ }\textbf {\bibinfo {volume} {111}},\ \bibinfo {pages}
  {247002} (\bibinfo {year} {2013})}\BibitemShut {NoStop}%
\bibitem [{\citenamefont {Hofer}\ \emph {et~al.}(2017)\citenamefont {Hofer},
  \citenamefont {Perarnau-Llobet}, \citenamefont {Miranda}, \citenamefont
  {Haack}, \citenamefont {Silva}, \citenamefont {Brask},\ and\ \citenamefont
  {Brunner}}]{hofer:2017}%
  \BibitemOpen
  \bibfield  {author} {\bibinfo {author} {\bibfnamefont {P.~P.}\ \bibnamefont
  {Hofer}}, \bibinfo {author} {\bibfnamefont {M.}~\bibnamefont
  {Perarnau-Llobet}}, \bibinfo {author} {\bibfnamefont {L.~D.~M.}\ \bibnamefont
  {Miranda}}, \bibinfo {author} {\bibfnamefont {G.}~\bibnamefont {Haack}},
  \bibinfo {author} {\bibfnamefont {R.}~\bibnamefont {Silva}}, \bibinfo
  {author} {\bibfnamefont {J.~B.}\ \bibnamefont {Brask}}, \ and\ \bibinfo
  {author} {\bibfnamefont {N.}~\bibnamefont {Brunner}},\ }\href@noop {}
  {\enquote {\bibinfo {title} {Markovian master equations for quantum thermal
  machines: local vs global approach},}\ } (\bibinfo {year} {2017}),\ \Eprint
  {http://arxiv.org/abs/1707.09211} {arXiv:1707.09211 [quant-ph]} \BibitemShut
  {NoStop}%
\bibitem [{\citenamefont {Gonz\'alez}\ \emph {et~al.}(2017)\citenamefont
  {Gonz\'alez}, \citenamefont {Correa}, \citenamefont {Nocerino}, \citenamefont
  {Palao}, \citenamefont {Alonso},\ and\ \citenamefont
  {Adesso}}]{gonzalez:2017}%
  \BibitemOpen
  \bibfield  {author} {\bibinfo {author} {\bibfnamefont {J.~O.}\ \bibnamefont
  {Gonz\'alez}}, \bibinfo {author} {\bibfnamefont {L.~A.}\ \bibnamefont
  {Correa}}, \bibinfo {author} {\bibfnamefont {G.}~\bibnamefont {Nocerino}},
  \bibinfo {author} {\bibfnamefont {J.~P.}\ \bibnamefont {Palao}}, \bibinfo
  {author} {\bibfnamefont {D.}~\bibnamefont {Alonso}}, \ and\ \bibinfo {author}
  {\bibfnamefont {G.}~\bibnamefont {Adesso}},\ }\href@noop {} {\enquote
  {\bibinfo {title} {Testing the validity of the local and global {G}{K}{L}{S}
  master equations on an exactly solvable model},}\ } (\bibinfo {year}
  {2017}),\ \Eprint {http://arxiv.org/abs/1707.09228} {arXiv:1707.09228
  [quant-ph]} \BibitemShut {NoStop}%
\bibitem [{\citenamefont {Alicki}\ \emph {et~al.}(2006)\citenamefont {Alicki},
  \citenamefont {Lidar},\ and\ \citenamefont {Zanardi}}]{alicki:2006}%
  \BibitemOpen
  \bibfield  {author} {\bibinfo {author} {\bibfnamefont {R.}~\bibnamefont
  {Alicki}}, \bibinfo {author} {\bibfnamefont {D.~A.}\ \bibnamefont {Lidar}}, \
  and\ \bibinfo {author} {\bibfnamefont {P.}~\bibnamefont {Zanardi}},\
  }\bibfield  {title} {\enquote {\bibinfo {title} {Internal consistency of
  fault-tolerant quantum error correction in light of rigorous derivations of
  the quantum {M}arkovian limit},}\ }\href {\doibase
  10.1103/PhysRevA.73.052311} {\bibfield  {journal} {\bibinfo  {journal} {Phys.
  Rev. A}\ }\textbf {\bibinfo {volume} {73}},\ \bibinfo {pages} {052311}
  (\bibinfo {year} {2006})}\BibitemShut {NoStop}%
\bibitem [{\citenamefont {Levy}\ \emph {et~al.}(2012)\citenamefont {Levy},
  \citenamefont {Alicki},\ and\ \citenamefont {Kosloff}}]{levy:2012pre}%
  \BibitemOpen
  \bibfield  {author} {\bibinfo {author} {\bibfnamefont {A.}~\bibnamefont
  {Levy}}, \bibinfo {author} {\bibfnamefont {R.}~\bibnamefont {Alicki}}, \ and\
  \bibinfo {author} {\bibfnamefont {R.}~\bibnamefont {Kosloff}},\ }\bibfield
  {title} {\enquote {\bibinfo {title} {Quantum refrigerators and the third law
  of thermodynamics},}\ }\href {\doibase 10.1103/PhysRevE.85.061126} {\bibfield
   {journal} {\bibinfo  {journal} {Phys. Rev. E}\ }\textbf {\bibinfo {volume}
  {85}},\ \bibinfo {pages} {061126} (\bibinfo {year} {2012})}\BibitemShut
  {NoStop}%
\bibitem [{\citenamefont {Szczygielski}\ \emph {et~al.}(2013)\citenamefont
  {Szczygielski}, \citenamefont {Gelbwaser-Klimovsky},\ and\ \citenamefont
  {Alicki}}]{szczygielski:2013}%
  \BibitemOpen
  \bibfield  {author} {\bibinfo {author} {\bibfnamefont {K.}~\bibnamefont
  {Szczygielski}}, \bibinfo {author} {\bibfnamefont {D.}~\bibnamefont
  {Gelbwaser-Klimovsky}}, \ and\ \bibinfo {author} {\bibfnamefont
  {R.}~\bibnamefont {Alicki}},\ }\bibfield  {title} {\enquote {\bibinfo {title}
  {Markovian master equation and thermodynamics of a two-level system in a
  strong laser field},}\ }\href {\doibase 10.1103/PhysRevE.87.012120}
  {\bibfield  {journal} {\bibinfo  {journal} {Phys. Rev. E}\ }\textbf {\bibinfo
  {volume} {87}},\ \bibinfo {pages} {012120} (\bibinfo {year}
  {2013})}\BibitemShut {NoStop}%
\bibitem [{\citenamefont {Holevo}(1982)}]{holevo:1982}%
  \BibitemOpen
  \bibfield  {author} {\bibinfo {author} {\bibfnamefont {A.~S.}\ \bibnamefont
  {Holevo}},\ }\href@noop {} {\emph {\bibinfo {title} {Probabilistic and
  Statistical Aspect of Quantum Theory}}}\ (\bibinfo  {publisher}
  {North-Holland},\ \bibinfo {address} {Amsterdam},\ \bibinfo {year}
  {1982})\BibitemShut {NoStop}%
\bibitem [{\citenamefont {Braunstein}\ and\ \citenamefont
  {Caves}(1994)}]{braunstein:1994}%
  \BibitemOpen
  \bibfield  {author} {\bibinfo {author} {\bibfnamefont {S.~L.}\ \bibnamefont
  {Braunstein}}\ and\ \bibinfo {author} {\bibfnamefont {C.~M.}\ \bibnamefont
  {Caves}},\ }\bibfield  {title} {\enquote {\bibinfo {title} {Statistical
  distance and the geometry of quantum states},}\ }\href {\doibase
  10.1103/PhysRevLett.72.3439} {\bibfield  {journal} {\bibinfo  {journal}
  {Phys. Rev. Lett.}\ }\textbf {\bibinfo {volume} {72}},\ \bibinfo {pages}
  {3439} (\bibinfo {year} {1994})}\BibitemShut {NoStop}%
\bibitem [{\citenamefont {Campbell}\ \emph {et~al.}(2017)\citenamefont
  {Campbell}, \citenamefont {Mehboudi}, \citenamefont {Chiara},\ and\
  \citenamefont {Paternostro}}]{campbell:2017}%
  \BibitemOpen
  \bibfield  {author} {\bibinfo {author} {\bibfnamefont {S.}~\bibnamefont
  {Campbell}}, \bibinfo {author} {\bibfnamefont {M.}~\bibnamefont {Mehboudi}},
  \bibinfo {author} {\bibfnamefont {G.~De}\ \bibnamefont {Chiara}}, \ and\
  \bibinfo {author} {\bibfnamefont {M.}~\bibnamefont {Paternostro}},\
  }\href@noop {} {\enquote {\bibinfo {title} {Global and local thermometry
  schemes in coupled quantum systems},}\ } (\bibinfo {year} {2017}),\ \Eprint
  {http://arxiv.org/abs/1705.01898} {arXiv:1705.01898 [quant-ph]} \BibitemShut
  {NoStop}%
\bibitem [{\citenamefont {Monras}(2013)}]{monras:2013}%
  \BibitemOpen
  \bibfield  {author} {\bibinfo {author} {\bibfnamefont {A.}~\bibnamefont
  {Monras}},\ }\href@noop {} {\enquote {\bibinfo {title} {Phase space formalism
  for quantum estimation of {G}aussian states},}\ } (\bibinfo {year} {2013}),\
  \Eprint {http://arxiv.org/abs/1303.3682} {arXiv:1303.3682 [quant-ph]}
  \BibitemShut {NoStop}%
\bibitem [{\citenamefont {Jiang}(2014)}]{jiang:2014}%
  \BibitemOpen
  \bibfield  {author} {\bibinfo {author} {\bibfnamefont {Z.}~\bibnamefont
  {Jiang}},\ }\bibfield  {title} {\enquote {\bibinfo {title} {Quantum {F}isher
  information for states in exponential form},}\ }\href {\doibase
  10.1103/PhysRevA.89.032128} {\bibfield  {journal} {\bibinfo  {journal} {Phys.
  Rev. A}\ }\textbf {\bibinfo {volume} {89}},\ \bibinfo {pages} {032128}
  (\bibinfo {year} {2014})}\BibitemShut {NoStop}%
\bibitem [{\citenamefont {\ifmmode~\check{S}\else \v{S}\fi{}afr\'anek}\ and\
  \citenamefont {Fuentes}(2016)}]{safranek:2016}%
  \BibitemOpen
  \bibfield  {author} {\bibinfo {author} {\bibfnamefont {D.}~\bibnamefont
  {\ifmmode~\check{S}\else \v{S}\fi{}afr\'anek}}\ and\ \bibinfo {author}
  {\bibfnamefont {I.}~\bibnamefont {Fuentes}},\ }\bibfield  {title} {\enquote
  {\bibinfo {title} {Optimal probe states for the estimation of {G}aussian
  unitary channels},}\ }\href {\doibase 10.1103/PhysRevA.94.062313} {\bibfield
  {journal} {\bibinfo  {journal} {Phys. Rev. A}\ }\textbf {\bibinfo {volume}
  {94}},\ \bibinfo {pages} {062313} (\bibinfo {year} {2016})}\BibitemShut
  {NoStop}%
\bibitem [{\citenamefont {Sparaciari}\ \emph {et~al.}(2016)\citenamefont
  {Sparaciari}, \citenamefont {Olivares},\ and\ \citenamefont
  {Paris}}]{sparaciari:2016}%
  \BibitemOpen
  \bibfield  {author} {\bibinfo {author} {\bibfnamefont {C.}~\bibnamefont
  {Sparaciari}}, \bibinfo {author} {\bibfnamefont {S.}~\bibnamefont
  {Olivares}}, \ and\ \bibinfo {author} {\bibfnamefont {M.~G.~A.}\ \bibnamefont
  {Paris}},\ }\bibfield  {title} {\enquote {\bibinfo {title} {Gaussian-state
  interferometry with passive and active elements},}\ }\href {\doibase
  10.1103/PhysRevA.93.023810} {\bibfield  {journal} {\bibinfo  {journal} {Phys.
  Rev. A}\ }\textbf {\bibinfo {volume} {93}},\ \bibinfo {pages} {023810}
  (\bibinfo {year} {2016})}\BibitemShut {NoStop}%
\end{thebibliography}%

\clearpage
\widetext
\begin{center}
\textbf{\large Supplement: Quantum Thermal Machine as a Thermometer}
\end{center}
\setcounter{equation}{0}
\setcounter{figure}{0}
\setcounter{table}{0}
\setcounter{page}{1}
\makeatletter
\renewcommand{\theequation}{S\arabic{equation}}
\renewcommand{\thefigure}{S\arabic{figure}}

\section{Approximate model}

Here we consider the approximate model that is obtained from the master equation in the main text by the replacement $(E_J/2)\hat{A}_c\hat{A}_h\rightarrow g$, leading to the Hamiltonian and current operator
\begin{equation}
\label{eq:toymodel}
\hat{H}_T=g\left(\hat{a}_h^\dag \hat{a}_c+H.c.\right),\hspace{2cm}\hat{I}_T=i2eg\left(\hat{a}_c^\dag \hat{a}_h-H.c.\right),
\end{equation}
where $g$ is treated as a fitting parameter. Such a model has first been investigated as a heat engine in Ref.~\cite{kosloff:1984}. An analogous engine that is based on coupled qubits instead of harmonic oscillators was introduced in Ref.~\cite{brunner:2012}. The time-evolution of the harmonic oscillators is described by the local Lindblad master equation
\begin{equation}
\label{eq:mastertoy}
\frac{\partial \hat{\rho}}{\partial t}  =  - i[\hat{H}_T , \hat{\rho}] +\kappa_h(n_B^h+1)\mathcal{D}[\hat{a}_h]\hat{\rho}+\kappa_hn_B^h\mathcal{D}[\hat{a}_h^\dag]\hat{\rho}+\kappa_c(n_B^c+1)\mathcal{D}[\hat{a}_c]\hat{\rho}+\kappa_hn_B^c\mathcal{D}[\hat{a}_c^\dag]\hat{\rho},
\end{equation}
with
\begin{equation}
\mathcal{D}[\hat{A}]\hat{\rho}=\hat{A}\hat{\rho}\hat{A}^\dag-\frac{1}{2}\{\hat{A}^\dag\hat{A},\hat{\rho}\},
\end{equation}
and the Bose-Einstein distribution
\begin{equation}
n_B^\alpha=\frac{1}{e^{\frac{\Omega_\alpha}{k_BT_\alpha}}-1}.
\end{equation}

As usual, differential equations for averages of operators are obtained by 
\begin{equation}
\partial_t \langle \hat{A}\rangle = {\rm Tr}\{\hat{A}\partial_t\hat{\rho}\}.
\end{equation}
Here we are interested in steady state values, where $\partial_t \langle \hat{A}\rangle=0$. For the mean current, this yields
\begin{equation}
\label{eq:currtoy}
\langle \hat{I}_T\rangle=-2e\frac{4\kappa_h\kappa_c g^2\left(n_B^h-n_B^c\right)}{\left(\kappa_h+\kappa_c\right)\left(\kappa_h\kappa_c+4g^2\right)},
\end{equation}
which is plotted in Fig.~\ref{fig:implementation}\,(a). The uncertainty in the temperature measurement plotted in Fig.~\ref{fig:implementation}\,(b) is obtained using 
\begin{equation}
\left(\frac{\partial f}{\partial P}\right)^2(\Delta P)^2=\left(\frac{\partial\langle\hat{I}_T\rangle}{\partial T_c}\right)^{-2}(\Delta I)^2,
\end{equation}
and
\begin{equation}
\label{eq:curtcderinv}
\frac{\partial\langle\hat{I}_T\rangle}{\partial T_c}=  \frac{2\kappa_h\kappa_c g^2}{(\kappa_h + \kappa_c)(\kappa_h\kappa_c + 4g^2)} \frac{e\Omega_c}{k_BT_c^2} \sinh^{-2}\left(\frac{\Omega_c}{2k_BT_c}\right),
\end{equation}
which yields the error
\begin{equation}
\label{eq:errortoy}
\Delta T_c = \sqrt{\left[\frac{k_B T_c^2}{e\Omega_c}\sinh^2\left(\frac{\Omega_c}{2k_BT_c}\right)\frac{(\kappa_h+\kappa_c)(\kappa_c\kappa_h+4g^2)}{2\kappa_c\kappa_h g^2}\right]^2(\Delta I)^2+\left(\frac{\Omega_c}{\Omega_h}\right)^2(\Delta T_h)^2}.
\end{equation}

\subsection{Steady-state solution}

The steady state solution of the master equation in Eq.~\eqref{eq:mastertoy} is Gaussian because the equation is bi-linear in creation and annihilation operators. It is therefore completely characterised by the first and second order moments of the quadrature operators
\begin{equation}
\hat{x}_\alpha = \frac{1}{\sqrt{2}}(\hat{a}_\alpha + \hat{a}_\alpha^\dagger),\hspace{2cm}\hat{p}_\alpha = \frac{1}{i\sqrt{2}}(\hat{a}_\alpha - \hat{a}_\alpha^\dagger) .
\end{equation}
Furthermore, since no term in the master equation induces displacements, the state is centred in phase space, $\langle\hat{x}_\alpha\rangle=\langle\hat{p}_\alpha\rangle=0$, and is thus determined by the second moments alone. 

We note that because the evolution is not unitary, it is not possible to determine the second moments simply by solving the equations of motions for the quadrature operators (working in the Heisenberg picture). While for unitary evolution, the time evolution of a product of operators is equal to the product of the time-evolved operators -- e.g. $(\hat{x}_\alpha\hat{p}_\alpha)(t) = \hat{x}_\alpha(t) \hat{p}_\alpha(t)$ -- this is not true in general. Hence, we need to derive the equations of motion for each of the second moments separately. We consider the 10 second moments $\avg{\hat{x}_c^2}$, $\avg{\hat{x}_c \hat{p}_c}$, $\avg{\hat{p}_c^2}$, $\avg{\hat{x}_h^2}$, $\avg{\hat{x}_h \hat{p}_h}$, $\avg{\hat{p}_h^2}$, $\avg{\hat{x}_c \hat{x}_h}$, $\avg{\hat{x}_c \hat{p}_h}$, $\avg{\hat{p}_c \hat{x}_h}$, $\avg{\hat{p}_c \hat{p}_h}$ (the remaining six are determined by the commutation relations). From \eqref{eq:mastertoy}, we find
\begin{equation}
\begin{aligned}
\frac{\partial}{\partial t}\avg{\hat{x}_c^2} & = 2g \avg{\hat{x}_c \hat{p}_h} - \kappa_c \avg{\hat{x}_c^2} + \kappa_c (n_B^c + \frac{1}{2}) , \\
\frac{\partial}{\partial t}\avg{\hat{x}_c\hat{p}_c} & = -g(\avg{\hat{x}_c\hat{x}_h} - \avg{\hat{p}_c\hat{p}_h}) - \kappa_c \hat{x}_c \hat{p}_c + \frac{i}{2}\kappa_c , \\
\frac{\partial}{\partial t}\avg{\hat{p}_c^2} & = -2g \avg{\hat{p}_c \hat{x}_h} - \kappa_c \avg{\hat{p}_c^2} + \kappa_c (n_B^c+\frac{1}{2}) , \\
\frac{\partial}{\partial t}\avg{\hat{x}_h^2} & = 2g \avg{\hat{p}_c \hat{x}_h} - \kappa_h \avg{\hat{x}_h^2} + \kappa_h(n_B^h + \frac{1}{2}) , \\
\frac{\partial}{\partial t}\avg{\hat{x}_h\hat{p}_h} & = -g(\avg{\hat{x}_c\hat{x}_h} - \avg{\hat{p}_c\hat{p}_h}) - \kappa_h \avg{\hat{x}_h \hat{p}_h} + \frac{i}{2}\kappa_h , \\
\frac{\partial}{\partial t}\avg{\hat{p}_h^2} & = -2g \avg{\hat{x}_c \hat{p}_h} - \kappa_h \avg{\hat{p}_h^2} + \kappa_h (n_B^h+\frac{1}{2}) , \\
\frac{\partial}{\partial t}\avg{\hat{x}_c\hat{x}_h} & = g (\avg{\hat{x}_c\hat{p}_c} + \avg{\hat{x}_h\hat{p}_h} - i) - \frac{1}{2}(\kappa_c + \kappa_h) \avg{\hat{x}_c \hat{x}_h} , \\
\frac{\partial}{\partial t}\avg{\hat{x}_c\hat{p}_h} & = -g(\avg{\hat{x}_c^2} - \avg{\hat{p}_h^2}) -\frac{1}{2}(\kappa_c + \kappa_h) \avg{\hat{x}_c \hat{p}_h} , \\
\frac{\partial}{\partial t}\avg{\hat{p}_c\hat{x}_h} & = -g(\avg{\hat{x}_h^2} - \avg{\hat{p}_c^2}) -\frac{1}{2}(\kappa_c + \kappa_h) \avg{\hat{p}_c \hat{x}_h} , \\
\frac{\partial}{\partial t}\avg{\hat{p}_c\hat{p}_h} & = - g (\avg{\hat{x}_c\hat{p}_c} + \avg{\hat{x}_h\hat{p}_h} - i) - \frac{1}{2}(\kappa_c + \kappa_h) \avg{\hat{p}_c \hat{p}_h} .
\end{aligned}
\end{equation}
The steady state is found by setting all the derivatives to zero. We then get
\begin{equation}
\label{eq:steadystate}
\begin{aligned}
\langle \hat{x}_c^2 \rangle= &  \langle \hat{p}_c^2 \rangle= \left(n_B^c + \frac{1}{2}\right) + \frac{g^2 \kappa_h}{(\kappa_c+\kappa_h)(g^2+\kappa_c\kappa_h)}(n_B^h - n_B^c) , \\
\langle \hat{x}_h^2 \rangle =& \langle \hat{p}_h^2 \rangle = \left(n_B^h + \frac{1}{2}\right) + \frac{g^2 \kappa_c}{(\kappa_c+\kappa_h)(g^2+\kappa_c\kappa_h)}(n_B^c - n_B^h) , \\
\langle \hat{x}_c\hat{p}_h \rangle= &-\langle \hat{p}_c\hat{x}_h \rangle= \frac{g \kappa_c\kappa_h}{(\kappa_c+\kappa_h)(g^2+\kappa_c\kappa_h)}(n_B^h-n_B^c) , \\
\langle \hat{x}_c\hat{p}_c \rangle= &\langle \hat{x}_h\hat{p}_h \rangle = \frac{i}{2} ,\hspace{2cm}
\langle \hat{x}_c\hat{x}_h \rangle= \langle \hat{p}_c\hat{p}_h \rangle = 0 . \\
\end{aligned}
\end{equation}
We note that for $g=0$ this is indeed a product of thermal states at temperatures $T_c$ and $T_h$ as expected. Also note that at the Carnot point we have $n_B^c = n_B^h$, which also results in a product of thermal states.

\section{Computing the Fisher information}
\label{app:fisher}

From the first and second moments of the quadratures -- i.e. from the covariance matrix, and the displacement vector (if the latter is non-zero) -- it is possible to compute the QFI, with respect to a given parameter of interest (in our case this is the cold bath temperature $T_c$). Various works have considered this problem, e.g.~\cite{monras:2013,jiang:2014,safranek:2016,sparaciari:2016}. Here, we will follow the method of \cite{sparaciari:2016}.

We define a vector of the quadratures $\mathbf{r} = (\hat{x}_c,\hat{x}_h,\hat{p}_c,\hat{p}_h)$ (note that this ordering is not the same as in \cite{sparaciari:2016}, nevertheless their method still applies, see also \cite{jiang:2014}). We restrict ourselves to the case where the first moments vanish, and define the covariance matrix as
\begin{equation}
\label{eq.covmatrix}
\Gamma_{ij} = \frac{1}{2}\avg{\hat{r}_i \hat{r}_j + \hat{r}_j \hat{r}_i} .
\end{equation}
The canonical commutation relations are given by $[\hat{r}_i,\hat{r}_j] = i \Omega_{ij}$, where the symplectic matrix $\Omega$ with this ordering of the $r_i$ is
\begin{equation}
\Omega = \left(\begin{array}{c c c c}
0 & 0 & 1 & 0 \\
0 & 0 & 0 & 1 \\
-1 & 0 & 0 & 0 \\
0 & -1 & 0 & 0 
\end{array}\right) .
\end{equation}

To compute the quantum Fisher information following \citep{sparaciari:2016}, one needs to find a symplectic diagonalisation of the covariance matrix. That is, one needs to find  a transformation $S$ such that
\begin{equation}
S\Omega S^T = \Omega ,
\end{equation}
and
\begin{equation}
S\Gamma S^T = \Gamma_S ,
\end{equation}
where $\Gamma_S$ is diagonal. One can then proceed to define the matrix $\Phi_S$ by
\begin{equation}
\label{eq:phis}
(\Phi_S)_{ij} = \frac{(\Omega^T \Gamma_S \dot{\Gamma}_S \Gamma_S \Omega + \frac{1}{4} \dot{\Gamma}_S)_{ij}}{2\lambda_i^2\lambda_j^2 - \frac{1}{8}} ,
\end{equation}
where the dot denotes derivation with respect to $T_c$, and $\lambda_\alpha$ are the eigenvalues of $\Gamma$. One further defines $\Phi = S^{-1} \Phi_S (S^T)^{-1}$, and the QFI is finally given by
\begin{equation}
\label{eq.qfigeneral}
F_{T_c} = \Tr [ \Omega^T \dot{\Gamma} \Omega \Phi ] .
\end{equation}

The symplectic diagonalisation is the most involved step of computing the QFI. For the steady-state solution in Eqs.~\eqref{eq:steadystate}
\begin{equation}
\Gamma = \left(
\begin{array}{cccc}
 a & 0 & 0 & -c \\
 0 & b & c & 0 \\
 0 & c & a & 0 \\
 -c & 0 & 0 & b \\
\end{array}
\right) ,
\end{equation}
where
\begin{equation}
a = \frac{1}{2} (1+2 n_B^c - \nu \kappa_c) , \hspace{1cm} b = \frac{1}{2} (1+2 n_B^h + \nu \kappa_h) , \hspace{1cm} c = \frac{1}{4g}\kappa_c \kappa_h \nu ,
\end{equation}
and
\begin{equation}
\nu = \frac{8 g^2 (n_B^c-n_B^h)}{(\kappa_c+\kappa_h) \left(4 g^2+\kappa_c \kappa_h\right)} .
\end{equation}
Denoting the normalized eigenvectors of $\Gamma$ as row vectors by $v_i$, $i=1,2,3,4$, a symplectic transformation diagonalizing $\Gamma$ can be constructed as
\begin{equation}
S = \left(
\begin{array}{c}
\text{Im}(v_1) \\
\text{Im}(v_3) \\ 
\text{Re}(v_1) \\
\text{Re}(v_3) 
\end{array}
\right) ,
\end{equation}
where Re and Im denote real and imaginary parts respectively. One finds that
\begin{equation}
\Gamma_S = S \Gamma S^T = \left(
\begin{array}{cccc}
 a' & 0 & 0 & 0 \\
 0 & b' & 0 & 0 \\
 0 & 0 & a' & 0 \\
 0 & 0 & 0 & b' \\
\end{array}
\right) ,
\end{equation}
with
\begin{equation}
a' = \frac{1}{2} \left(a+b+\sqrt{(a-b)^2+4 c^2}\right) , \hspace{1cm} b' = \frac{1}{2} \left(a+b-\sqrt{(a-b)^2+4 c^2}\right) .
\end{equation}
From Eqs.~\eqref{eq:phis} and \eqref{eq.qfigeneral}, and by using that close to the Carnot point
\begin{equation}
n_B^c - n_B^h = \frac{1}{e^{\Omega_c/T_c}-1} - \frac{1}{e^{\Omega_c/T_c+\epsilon}-1} \approx {n}^c_B({n}^c_B+1) \epsilon,
\end{equation}
the expression for the QFI at the Carnot point can be found by taking the limit $\epsilon\rightarrow 0$ and reads
\begin{equation}
F_{T_c} = \frac{\kappa _c^2 \left(8 g^2 \kappa _h \left(\kappa _c+2 \kappa _h\right)+\kappa _h^2 \left(\kappa _c+\kappa _h\right){}^2+32 g^4\right) }{4  \left(\kappa _c+\kappa _h\right){}^2 \left(\kappa _c \kappa _h+4 g^2\right){}^2} \frac{\Omega _c^2}{T_c^4} \frac{1}{\sinh^2(\Omega_c/2T_c)} .
\end{equation}
The parameter $\beta$ in the main text thus reads
\begin{equation}
\beta = \frac{2 \left(\kappa _c+\kappa _h\right) \left(\kappa _c \kappa _h+4 g^2\right)}{\kappa _c \sqrt{8 g^2 \kappa _c \kappa _h+\kappa _h^2 \left(\kappa _c^2+16 g^2\right)+2 \kappa _c \kappa _h^3+32 g^4+\kappa _h^4}} .
\end{equation}

\section{Computing the uncertainty from a current measurement}
\label{app:current}

We aim to use Eq.~\eqref{eq:errorprop} for the simple model given in Eq.~\eqref{eq:toymodel} in the steady state. The mean current $\langle \hat I_T \rangle$ and its derivative with respect to $T_c$ are given in Eqs.~\eqref{eq:currtoy} and \eqref{eq:curtcderinv} respectively. The variance $(\Delta \hat I_T)^2 = \langle \hat I_T^2 \rangle - \langle \hat I_T \rangle^2$ reads

\begin{equation}
\begin{split}
(\Delta \hat I_T)^2 = (2e)^2 g^2 \bigg( n_B^c(n_B^h+1) + n_B^h(n_B^c+1) & + \frac{8 g^2(n_B^h-n_B^c)}{(\kappa_h+\kappa_c)(\kappa_h\kappa_c + 4g^2)} \left[\kappa_h(n_B^h + 1/2) - \kappa_c(n_B^c+1/2)\right] \\
& - \frac{32 g^4 \kappa_h\kappa_c(n_B^h-n_B^c)^2}{(\kappa_h+\kappa_c)^2(\kappa_h\kappa_c + 4g^2)^2} \bigg) ,
\end{split}
\end{equation}
which simplifies to
\begin{equation}
(\Delta \hat{I}_T)^2 = 2(2e)^2 g^2 n_B^c(n_B^c+1) = 2 e^2 g^2 \frac{1}{\sinh^2(\Omega_c/2T_c)} ,
\end{equation}
at the Carnot point where we have $n_B^c=n_B^h$.
Hence, we get that with a current measurement at the Carnot point
\begin{equation}
(\Delta T_c)^2_{\hat{I}_T} = \frac{(\kappa_h + \kappa_c)^2(\kappa_h\kappa_c + 4g^2)^2}{2\kappa_h^2\kappa_c^2g^2} \frac{T_c^4}{\Omega_c^2} \sinh^2(\Omega_c/2T_c) ,
\end{equation}
which results in
\begin{equation}
\alpha = \frac{(\kappa_h + \kappa_c)(\kappa_h\kappa_c + 4g^2)}{\sqrt{2}\kappa_h\kappa_cg} .
\end{equation}

{
\section{Heat leaks and spurious modes}
Heat leaks, connecting the hot and the cold bath, as well as spurious modes which couple to the Cooper pairs can lead to an efficiency that deviates from the Otto efficiency, impeding the measurement of the cold temperature. Here we discuss these issues in detail, showing that they do not pose a major problem.

In the absence of heat leaks, the oscillators thermalize to the bath temperatures whenever the back-action of the Cooper pair current can be neglected (this is the case at the Carnot point, or if the voltage is not at resonance $2eV\neq \Omega_h-\Omega_c$, or if either $E_J$ or $\lambda_\alpha$ go to zero). The thermometer then measures the temperature of the cold bath as discussed in the main text.
In the presence of heat leaks, the oscillators will not necessarily thermalize to the bath temperatures even when the back-action of the current can be neglected. Denoting the temperatures of the oscillators in the absence of a Cooper pair current as $T_\alpha^*$, the corresponding occupation numbers read $n_\alpha^*=[1-e^{\Omega_\alpha/k_BT_\alpha^*}]^{-1}$. Importantly, the power consumption of the engine will go to zero when $n_c^*=n_h^*$ which implies $\Omega_c/T_c^*=\Omega_h/T_h^*$. If the hot temperature measurement required for our proposal measures $T_h^*$, our thermometer will thus determine $T_c^*$.
Since this is the steady-state temperature of the oscillator with frequency $\Omega_c$ (in the absence of a current), this is arguably the temperature of interest. If the hot temperature measurement still measures the bath temperature $T_h$, any difference $T_h\neq T_h^*$ can be accounted for in the error $\Delta T_h$. We note that only heat leaks that involve the oscillator degrees of freedom will lead to $T_\alpha^*\neq T_\alpha$. Any additional heat currents in or out of the baths that do not involve the oscillator degrees of freedom do not change the results of the main text in any way. This is important since in practice the baths will be stabilized with heat flows that are not taken into account in the analysis.

Spurious modes which have a finite density of states at the Josephson frequency $\Omega_h-\Omega_c$ lead to a finite current $I_0$ (dissipating power) even when $\langle \hat{n}_c\rangle = \langle \hat{n}_h\rangle$. This could in principle be detrimental to our thermometry scheme. However, by sweeping the voltage, the environment that couples to the Josephson junction can be characterized fairly well. Experiments suggest that spurious modes at the Josephson frequency can be avoided \cite{hofheinz:2011, westig:2017}. Even if such modes are present, one can in principle determine $I_0$ if their presence is known. In this case, one can correct for the dissipative current.}

\end{document}